\documentclass[letterpaper,journal]{IEEEtran}

\usepackage{amsmath,amsfonts,amssymb,amsthm,mathtools,bm}
\usepackage{cite}

\usepackage{color}
\usepackage{xcolor}

\usepackage{graphicx}
\usepackage{epstopdf}
\usepackage{float}
\usepackage{stfloats}

\usepackage[caption=false,font=normalsize,labelfont=sf,textfont=sf]{subfig}
\usepackage[justification=centering]{caption}

\usepackage{booktabs}
\usepackage{multirow}
\usepackage{tabularx}
\usepackage{threeparttable}
\usepackage{verbatim}
\usepackage{setspace}
\usepackage[most]{tcolorbox}

\usepackage[linesnumbered,ruled,vlined]{algorithm2e}
\SetKwInput{KwInput}{Input}
\SetKwInput{KwOutput}{Output}
\SetKwBlock{Where}{where}{}
\SetKwComment{tcp}{// }{}
\DontPrintSemicolon  

\usepackage[hyphens]{url} 
\usepackage[ breaklinks=true]{hyperref} 

\allowdisplaybreaks[0] 

\usepackage[switch]{lineno}

\usepackage{fancyhdr}

\begin{document}

\title{Enhanced Velocity-Adaptive Scheme: Joint Fair Access and Age of Information Optimization in Vehicular Networks}

\author{Xiao Xu, Qiong Wu,~\IEEEmembership{Senior Member,~IEEE}, Pingyi Fan,~\IEEEmembership{Senior Member,~IEEE}, \\
Kezhi Wang,~\IEEEmembership{Senior Member,~IEEE}, Nan Cheng,~\IEEEmembership{Senior Member,~IEEE},\\ Wen Chen,~\IEEEmembership{Senior Member,~IEEE}, 
and Khaled B. Letaief,~\IEEEmembership{Fellow,~IEEE}

\thanks{Part of this paper has been accepted by IEEE RFAT 2025 conference. This work was supported in part by Jiangxi Province Science and Technology Development Programme under Grant No. 20242BCC32016, in part by the National Natural Science Foundation of China under Grant No. 61701197, in part by the National Key Research and Development Program of China under Grant No. 2021YFA1000500(4), in part by the Shanghai Kewei under Grant 22JC1404000and Grant 24DP1500500, in part by the Research Grants Council under the Areas of Excellence Scheme under Grant AoE/E-601/22-R and in part by the 111 Project under Grant No. B23008. (Corresponding author: Qiong Wu.)

Xiao Xu and  Qiong Wu are with the School of Internet of Things Engineering, Jiangnan University, Wuxi 214122, China, and also with the School of Information Engineering, Jiangxi Provincial Key Laboratory of Advanced Signal Processing and Intelligent Communications, Nanchang University, Nanchang 330031, China (e-mail:  xuxiao@stu.jiangnan.edu.cn, qiongwu@jiangnan.edu.cn). 


Pingyi Fan is with the Department of Electronic Engineering, State Key laboratory of Space Network and Communications,Beijing National Research Center for Information Science and Technology, Tsinghua University, Beijing 100084, China (email: fpy@tsinghua.edu.cn).

Kezhi Wang is with the Department of Computer Science, Brunel University, London, Middlesex UB8 3PH, U.K. (email: Kezhi.Wang@brunel.ac.uk)

Nan Cheng is with the State Key Laboratory of ISN and the School of Telecommunications Engineering, Xidian University, Xi’an 710071, China (e-mail: dr.nan.cheng@ieee.org).

Wen Chen is with the Department of Electronic Engineering, Shanghai Jiao Tong University, Shanghai 200240, China (e-mail: wenchen@sjtu.edu.cn).

Khaled B. Letaief is with the Department of Electrical and Computer Engineering, the Hong Kong University of Science and Technology, Hong Kong (e-mail: eekhaled@ust.hk).

}
}


\maketitle

\begin{abstract}

In this paper, we consider the fair access problem and the Age of Information (AoI) under 5G New Radio (NR) Vehicle-to-Infrastructure (V2I) Mode 2 in vehicular networks. Specifically, vehicles follow Mode 2 to communicate with Roadside Units (RSUs) to obtain accurate data for driving assistance. Nevertheless, vehicles often have different velocity when they are moving in adjacent lanes, leading to difference in RSU dwell time and communication duration. This results in unfair access to network resources, potentially influencing driving safety.  
To ensure the freshness of received data, the AoI should be analyzed. Mode 2 introduces a novel preemption mechanism, necessitating simultaneous optimization of fair access and AoI to guarantee timely and relevant data delivery. We propose a joint optimization framework for vehicular network, defining a fairness index and employing Stochastic Hybrid Systems (SHS) to model AoI under preemption mechanism. By adaptively adjusting the selection window of Semi-Persistent Scheduling (SPS) in Mode 2, we address the optimization of fairness and AoI. We apply a large language model (LLM)-Based Multi-objective Evolutionary Algorithm Based on Decomposition (MOEA/D) to solve this problem. Simulation results demonstrate the effectiveness of our scheme in balancing fair access and minimizing AoI.
\end{abstract}


\IEEEpeerreviewmaketitle

\section{Introduction}
\label{sec1}
\IEEEPARstart{A}{utonomous} driving technology represents a promising innovation that is expected to transform transportation systems and serve as a critical component of future smart cities \cite{32}. Many companies, such as Baidu Apollo, BYD, and Tesla, are currently developing autonomous driving solutions, with pilot autonomous vehicles already operating in cities like Beijing, San Francisco, Shanghai, and Los Angeles \cite{33}.  
Information acquisition is a critical component of autonomous driving technology. To perceive their surroundings, vehicles are typically equipped with multiple sensors, such as high-definition cameras and Light Detection And Ranging (LiDAR) \cite{34,57}. However, the large amount of redundant information imposes significant computational burdens on systems.  

Cloud computing has been widely adopted to address computational resource limitations \cite{35,59,58,104,111}. Specifically, vehicles upload raw data to remote cloud servers with high processing power, which return refined information after computation. However, the geographical distance between cloud servers and vehicles introduces substantial transmission latency, rendering this approach unsuitable for high-speed vehicular environments \cite{36,105}. This issue is addressed by using the Edge computing framework which uses the edge server to provide low latency information to vehicles \cite{56}. Specifically, with the 3rd Generation Partnership Project (3GPP) Release 16 establishing the first 5G New Radio (NR) standard, two communication modes are defined: Mode 1 and Mode 2 \cite{49}. In Mode 1, Base Stations (BS) allocate communication resources, requiring vehicles to move within network coverage. This not only increases latency but also restricts mobility. Conversely, Mode 2 enables vehicles to independently select Sidelink (SL) resources, allowing communication even outside network \cite{50}. 
In addition, in high-speed scenarios, vehicles frequently enter and exit the coverage area of the base station, which may result in instantaneous communication interruptions. Frequent coverage switching can also cause occasional high latency. Furthermore, in remote areas, there are regions where the base station under Mode 1 is difficult to cover. Therefore, compared with Mode 2, Mode 1 can lead to situations that are relatively dangerous for high-speed vehicles. In contrast, Mode 2 can autonomously select communication resources based on semi-persistent scheduling (SPS), allowing independent decision-making without network coverage and avoiding the scheduling time required for communication with the base station, thereby enabling low-latency communication.
Thus, vehicles typically adopt the SPS mechanism in Mode 2 for communication \cite{1,107,108,109}. 
However, the SPS mechanism introduces new challenges:  vehicles in adjacent lanes with varying speeds experience unequal RSU dwell times,  leading to unfair access and degraded AoI—issues not fully addressed by prior works.

Under the SPS mechanism, vehicles first reserve SL resources and then upload redundant data to Roadside Units (RSUs) equipped with edge servers \cite{46,60,103,110}. These edge servers help to process the data and return actionable information to vehicles. Since RSUs are deployed near roadways, the communication latency remains sufficiently low to support real-time driving decisions, effectively mitigating safety risks caused by delayed updates.
However, compared with LTE Mode 4, 5G NR V2X Mode 2 introduces a unique preemption mechanism, whereby high-priority vehicles can occupy the communication resources of low-priority vehicles to achieve prioritized communication—this is a mechanism that does not exist in LTE Mode 4. Therefore, although the introduction of this mechanism ensures that higher-priority vehicles can transmit first, once high-priority vehicles frequently preempt resources, it becomes difficult for low-priority vehicles to communicate. This not only exacerbates the problem of fair access but also leads to an increase in the AoI of low-priority vehicles.

Specifically, on highways, vehicles in different lanes operate at varying speeds. This results in unequal dwell times within RSU coverage: faster vehicles communicate with RSUs for shorter durations, receiving less information compared to slower vehicles. 
Specifically, within the same RSU coverage area, the amount of information received by each vehicle should ideally be equal. However, in reality, faster-moving vehicles often obtain less information due to shorter communication times with the RSU. This may result in situations where only slower vehicles receive information that should have been available to all vehicles within the RSU's coverage, or where high-speed vehicles fail to timely obtain the requested information. Such unfair information acquisition within the same area can pose safety risks to nearby vehicles, especially those moving at high speeds. Therefore, it is necessary to define a fairness index to ensure fair access for all vehicles within the same region.
This unfair access in 5G NR V2I Mode 2 compromises the decision-making accuracy and safety of high-speed vehicles. 
Additionally, data freshness, measured using the Age of Information (AoI) \cite{44,102,61}, critically impacts safety in high-speed vehicular networks. Elevated AoI delays critical updates, hindering vehicles’ ability to respond to dynamic environments. 

However, considering the unique preemption mechanism in 5G NR V2X Mode 2, high-priority vehicles can preempt the communication resources of low-priority vehicles. Admittedly, this grants high-priority vehicles the privilege of prioritized communication, ensuring the communication speed required for emergency situations. Nevertheless, if high-priority vehicles repeatedly preempt the resources of low-priority vehicles, it can become difficult for low-priority vehicles to obtain communication resources, significantly exacerbating the fairness issue. At the same time, the AoI of low-priority vehicles increases as they wait for resources. Therefore, if fairness in access is considered in isolation under the preemption mechanism, it may lead to an overall increase in network latency while further intensifying fairness problems. Conversely, if only the overall AoI of the network is considered, fairness in access is difficult to guarantee. Consequently, designing a scheme that can jointly optimize fair access and AoI under 5G NR V2X Mode 2, especially considering the existence of the preemption mechanism, is of critical importance. Moreover, in our previous research \cite{63}, we only considered fair access and did not take into account the impact of the preemption mechanism on fairness and AoI. Currently, there is no solution that comprehensively considers both fair access and AoI under the preemption mechanism of 5G NR, which has motivated us to undertake this work. 

This paper proposes a multi-objective optimization framework for 5G NR V2X Mode 2 vehicular networks, where the selection window size is adaptively adjusted based on vehicle speed to ensure fair access and minimize AoI. The main contributions are outlined below\footnote{Source code can be found at : \url{https://github.com/qiongwu86/Enhanced-Velocity-Adaptive-Scheme-Joint-Fair-Access-and-Age-of-Information-Optimization-in-iov}}:

\begin{itemize}
\item[1)]  We defined a fairness index under 5G NR V2X Mode 2, which takes into account the SPS scheduling and the resource collision probability in Mode 2. Based on this, we adaptively adjust the selection window size according to vehicle speed to achieve fair access for vehicles with different speeds.
\item[2)] Considering that the unique preemption mechanism in 5G NR V2X Mode 2 aggravates the fair access problem among vehicles with different speeds and leads to an increase in the AoI of vehicles \cite{64}, we adopted a stochastic hybrid system (SHS) to model the AoI of vehicles. This modeling specifically takes into account the unique preemption mechanism in 5G NR, and ultimately establishes the relationship between the vehicles’ AoI and the selection window size.
\item[3)]  We proposed a multi-objective optimization scheme that jointly considers fairness and AoI, aiming to simultaneously optimize fair access and the potential increase in AoI caused by the change of selection window under the preemption mechanism in order to realize fair access.
 A Large Language Model (LLM)-Based Multi-objective Evolutionary Algorithm Based on Decomposition (MOEA/D) is applied, deriving optimal window sizes for different vehicles. Our results demonstrate that the LLM-Based algorithm outperforms traditional methods in convergence speed and solution diversity. Simulation experiments validate the effectiveness of our approach in achieving both fairness and AoI minimization.  

\end{itemize}

The following structure is arranged as below:
Section \ref{sec2} discusses related work. Section \ref{sec3} introduces the system model. Section \ref{sec4} presents the fairness metric. Section \ref{sec5} analyzes the average AoI in networks using SHS. Section \ref{sec6} defines the optimization problem and details the LLM-Based MOEA/D. Section \ref{sec7} presents simulation results. Finally, Section \ref{sec8} summarizes the findings of the paper.



\section{Related Works}
\label{sec2}
This section provides an overview of related studies.
\subsection{SPS mechanism}
Existing works explored SPS mechanisms in C-V2X\cite{26,27,28}. 
In \cite{26}, Amr \textit{et al.} developed the C-V2X simulator analyzing the impact of resource pool configurations and essential parameters. 
In \cite{27}, Ye \textit{et al.} introduced an innovative scheme and corresponding scheduler which uses past transmission data to decrease delay. In \cite{28}, Gu \textit{et al.} developed an analytical SPS model to quantify the effects of beacon rate and configurations on access collision rate and latency outage rate. This model provides critical insights for optimizing communication configurations, comprising signal detection radius, transmission power and resource reservation. The optimized system maintains consistent packet delivery rates and low delays under varying traffic density conditions.

Additional research has focused on SPS mechanisms in 5G NR-V2X \cite{29,30,31}. 
In \cite{29}, Malik \textit{et al.} proposed an enhanced SPS scheme for aperiodic traffic resource reservation. This approach dynamically adjusts the sensing window size according to traffic distribution intensity and vehicle velocity while incorporating a re-evaluation mechanism to confirm resource availability through repeated sensing after selection. 
In \cite{30}, Daw \textit{et al.} presented a method considering data priority that classifies urgent vehicles as High-Priority (HP) with elevated Reselection Counter (RC), enabling them to transmit CAM messages on reserved channel resources. This reduces collision probability among heterogeneous vehicles. 
In \cite{31}, Luca \textit{et al.} analyzed and compared the effectiveness of SPS with Dynamic Scheduling (DS) across varying data flows and Packet Delay Budget (PDB) limits. Their results show that adaptive scheduling strategies, allowing vehicles to choose the optimal method for traffic patterns, achieve superior performance in hybrid scenarios combining periodic and aperiodic traffic. However, none of these studies address the joint optimization of fairness and AoI.
\subsection{Fairness of network}  
Several studies have focused on ensuring fairness in wireless networks \cite{38,39,40}. 
In \cite{38}, Park \textit{et al.} proposed a power control-based equitable channel access scheme for wireless networks. This scheme employs a distributed channel access algorithm that adjusts individual node access probabilities to achieve fairness in the update intervals of randomly accessed network states. Simulation results demonstrated high information coverage while maintaining fairness among nodes.
In \cite{39}, Zhang \textit{et al.} introduced a dynamic MAC protocol to address temporal unfairness in dynamic channel access. Their approach assigns different time slot allocation schemes to different nodes, thereby ensuring spatial access fairness. Similarly, in \cite{40}, Gibson \textit{et al.} modeled fairness criteria for MAC protocols in multi-hop sensor networks, ensuring that transmission rates between sensors and BS remain equal.

Additionally, some studies have investigated fairness in vehicle access scenarios. 
In \cite{9}, Wan \textit{et al.} considered the unfair access issue between vehicles and RSUs under the IEEE 802.11p protocol. They defined a fairness index and proposed an approach that adjusts the minimum contention window according to vehicle speeds to realize fair access. Furthermore, they modeled and optimized the network’s AoI.
In \cite{41}, Muhammed \textit{et al.} addressed the issue of imbalanced resource allocation due to significant variations in vehicle density across different regions. They proposed a scheme for fair resource allocation among edge nodes, which was evaluated in real-world scenarios.
In \cite{42}, Wang \textit{et al.} examined the unfair network resource allocation problem caused by structural asymmetry in uplink and downlink connections. They proposed an edge network system that leverages edge computing to provide fair access services. 
In\cite{43}, Harigovindan \textit{et al.} studied fair vehicle access in V2I networks under the IEEE 802.11p protocol. Their approach accounts for unfair access caused by variations in vehicle speeds and derives the optimal Contention Window (CW) to ensure fairness.

In our previous work \cite{63}, we considered the issue of fair access in 5G NR V2X. However, that work only addressed the difference between 5G NR V2X and LTE V2X in terms of resource collision probability and did not take into account the unique preemption mechanism introduced in 5G NR V2X. This mechanism can result in high-priority vehicles frequently occupying the communication resources of low-priority vehicles, thereby exacerbating the problem of fair access and increasing the AoI of low-priority vehicles as they wait for resources. This situation can be particularly dangerous in high-speed vehicular environments. Therefore, considering only fair access is one-sided after the introduction of the preemption mechanism in 5G NR, which may cause the increase of AoI due to the change of selection window \cite{101}. 

However, in \cite{9}, Wan \textit{et al.} did jointly consider access fairness and AoI. However, this work only focused on fair access under the IEEE 802.11p protocol and did not consider the novel 5G NR V2X protocol. In addition, their AoI modeling did not account for the presence of the preemption mechanism.

Moreover, to the best of our knowledge, current research on fairness does not take into account the 5G NR V2X protocol, especially the preemption mechanism. Moreover, there is no comprehensive work that optimizes both vehicle fair access and AoI after the introduction of the preemption mechanism in 5G NR. Additionally, considering the preemption mechanism in 5G NR and deriving AoI using SHS seems to be an unexplored area, which motivates us to undertake this work.

\subsection{Age of information}  
As a metric for data freshness, AoI is indispensable in high-speed scenarios, especially for vehicular networks. Consequently, extensive research has focused on AoI optimization. 
In \cite{20}, Azizi \textit{et al.} used reinforcement learning in C-V2X to minimize AoI while maximizing energy efficiency, and investigated the impact of increased inter-vehicle spacing on AoI. 
In \cite{21}, Zhang \textit{et al.} analyzed the relationship between multi-priority queues and Non-Orthogonal Multiple Access (NOMA) with AoI, proposing a DRL-based method to optimize both energy consumption and AoI. 
In \cite{22}, Ali \textit{et al.} modeled AoI using stochastic hybrid systems in CSMA environments, minimizing average AoI by calibrating backoff times. 
In \cite{23}, Roy \textit{et al.} extended stochastic hybrid systems to derive generalized AoI results applicable to multi-source systems, reducing AoI evaluation complexity to stationary distribution analysis of finite-state Markov chains.
In \cite{62}, Yu \textit{et al.} considered the problem of detachment caused by failures in an edge-enabled vehicular metaverse, which disrupts the sense of immersion. They proposed a scheme based on redundant backups and maintaining the AoI of the backups to avoid such detachment, with the focus primarily on preventing disconnection. However, this work did not take into account the issue of fair access resulting from vehicles with different speeds generating different amounts of information exchange. Moreover, it also did not consider modeling the AoI using SHS.

For 5G NR V2X, Liu \textit{et al.} \cite{24} analyzed SPS parameters in Mode 2 to ensure message freshness. 
In \cite{25}, Saad \textit{et al.} developed a deep reinforcement learning-based congestion control mechanism addressing Mode 2 NR-V2X's inefficiency in handling aperiodic packet scheduling while optimizing AoI. However, these studies neglect the fair access in the system.

Existing research lacks solutions that jointly optimize fair access and AoI, motivating our work to address this gap.

\section{System Model}
\label{sec3}
This section presents the system model, which is primarily divided into the scenario model and the SPS protocol model under 5G NR V2X Mode 2.
\begin{figure*}[htbp]
	\centering
	\includegraphics[width=\linewidth, scale=1.00]{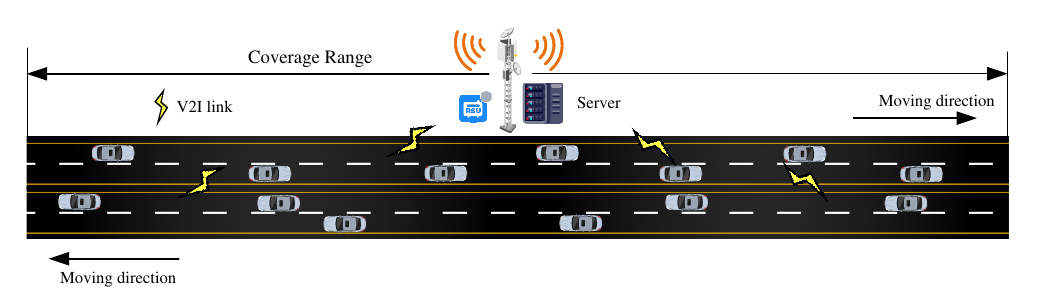}
	\caption{Scenario Model}
	\label{fig1}
	\vspace{-0.7cm}
\end{figure*}

\subsection{Scenario}
Fig. 1 presents our scenario model, where we consider a multi-lane highway scenario within the RSU's range, which is deployed along the roadside. The RSU is equipped with an edge server that has sufficient computational resources. It is assumed that vehicles attempt to communicate with the RSU to obtain useful information whenever they are within its coverage area. The vehicles in the same lane move at a uniform speed, while vehicles in adjacent lanes differ in velocity at least 4 m/s. Each vehicle communicates with the RSU and receives useful information. Considering that the downlink data volume is significantly larger than the uplink data volume, we focus only on the uplink transmission \cite{18}.

\subsection{SPS }
Following the Mode 2 SPS protocol, vehicles autonomously allocate communication resources instead of relying on base station assignments, allowing resource allocation even in the absence of network coverage. Specifically, as shown in Fig. 2, the channel is divided into different subchannels. These subchannels are utilized for both data transmission and control information. Data is transmitted in the form of Transport Blocks (TBs), each of which carries a SideLink Control Information (SCI) which provides metadata about TB. When the RC reaches 0, vehicle need to choose new resources.

Each time a vehicle selects resources, it follows a sensing-based mechanism, with a Resource Reservation Interval (RRI) between consecutive resource selections. The vehicle first identifies candidate resources within a selection window, whose size is determined by the vehicle’s application requirements. Next, the vehicle discards certain resources according to the following conditions:
\begin{itemize}
	\item[1)] Resources reserved by other vehicles are excluded.
	\item[2)] Resources with a time-averaged Reference Signal Received Power (RSRP) exceeding a threshold are excluded.
\end{itemize}
\begin{figure}[htbp]
	\centering
	\includegraphics[width=\linewidth, scale=1.00]{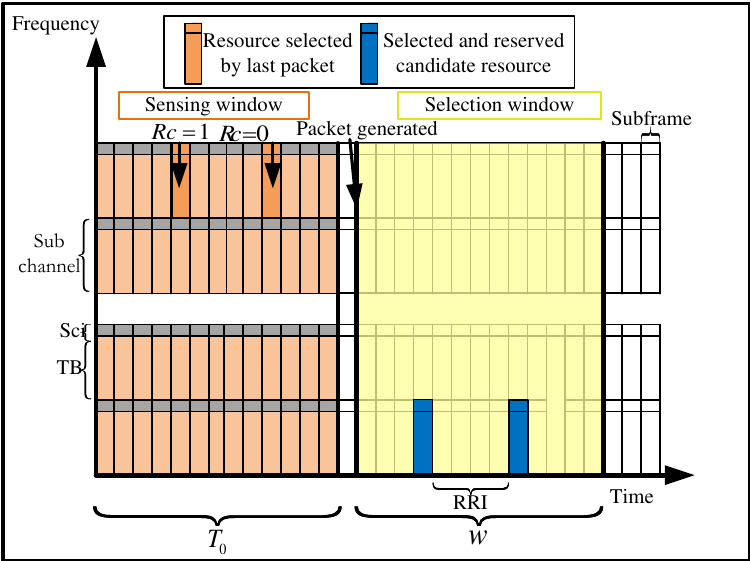}
	\caption{SPS Model}
	\label{fig2}
\end{figure}

After eliminating unsuitable resources, from the available resources, the vehicle chooses one at random, further reducing the probability of collisions with other vehicles. Once a resource is selected, the vehicle proceeds with TB transmission.

Compared to LTE Mode 4, Mode 2 introduces a preemption mechanism, enabling more flexible resource allocation based on traffic priority. If a priority threshold is predefined, a vehicle will release its resources when another vehicle with a higher priority exceeds this threshold. This mechanism helps prevent resource collisions with high-priority transmissions. Specifically, if a lower-priority vehicle has already reserved a resource within its selection window, and a higher-priority vehicle subsequently requires the same resource, the lower-priority vehicle will relinquish its reservation, allowing the higher-priority vehicle to use the resource.

\section{Fairness Index}
\label{sec4}

This section establishes the link between fairness, selection window size, and vehicle velocity. First, we propose a fairness index to measure fair access. Then, we analyze the successful transmission probability of vehicle packets. Finally, we find that the fairness index relates to vehicle velocity and window size. Table I lists the parameters used in this paper.

\begin{table}\footnotesize
	
	\caption{Symbols in this paper }
	\label{tab1}
	\centering
	\begin{tabular}{|c|p{6.6cm}|}
		\hline
		\textbf{Symbol} &\textbf{Description}\\
		\hline
		$Bit^{i}$  & \multicolumn{1}{m{6.6cm}|}{ The expected total data size transmitted by vehicle $i$ during the coverage area of the RSU.}\\
		\hline
		$T_i$  &  The time vehicle $i$ spends within  RSU's range.\\
		\hline
		$P_{PRR^i}$  &  The probability of successful transmission\\
		\hline
		$R$ & The coverage range of the RSU. \\
		\hline
		$C_i$ & Vehicle $i$'s bit rate. \\
		\hline
		$B$ & The bandwidth. \\
		\hline
		${p}_{i}$ &  The transmission power of vehicle $i$. \\
	    \hline
	
		${h}_{i}$   &The channel gain. \\
		\hline
		${d}_{i}$  &The geometric distance between vehicle $i$ and the RSU. \\
		\hline
		${P}_{o}^{i}$ & The position of vehicle $i$. \\
		\hline
		$f_{d}^{i}$ &The Doppler frequency of vehicle $i$. \\
		\hline
		
		$K_{index}^{i}$ &The fairness index of vehicle $i$. \\
		\hline
		$K_{index}$ &The overall fairness of the network \\
		\hline
		$q(t)$ &The channel state. \\
		\hline
		\( x_0(t) \) & The AoI at the RSU.  \\
		\hline
		\( x_1(t) \) & The AoI at vehicle $k$  \\
		\hline
		\( A_L \)  &The reset mapping for that transition. \\
		\hline
		\( x' \) & The continuous process values before and after the reset \\
		\hline
		\( \lambda \)  & The transition rate of the discrete process \( L \). \\
		\hline
		\( v_{q_0} \) &The stationary correlation between \( q_l \) and \( x_0 \). \\
		\hline
		\( v_{q_1} \) &The stationary correlation between \( q_l \) and \( x_1 \). \\
		\hline
		$\bar{\Delta}_k$ &The averaged AoI for link \( k \).  \\
		\hline
		
		\( H_k \)  & The average successful transition rate of link \( k \).\\
		\hline
		\( R_k \) &The averaged failed transition rate. \\
		\hline
		
		${\bar{\pi}}_{\bar{q}}$ & The steady-state distribution of \(q\). \\
		\hline
		\( T_s \)  &The average transmission success time. \\
		\hline
		\( T_{ini} \) & The total transmission time. \\
		\hline
		\( t_r \)  &The time required for retransmission. \\
		\hline
		\( T_{sch} \) & The time required for resource scheduling. \\
		\hline
		\( T_{pkt} \) &The actual transmission time. \\
		\hline
		\( t_p \) & The time required for the sender to deal the data.\\
		\hline
		\( t_{fa} \) & Constrained by the duration of the time slot. \\
		\hline
		\( T_w \) &The time required for resource scheduling. \\
		\hline
		$t_{\text{NACK}}$ & The transmission delay of the NACK. \\
		\hline
		\( T_f \) & The average transmission failure time. \\
		\hline
		${N}_{Sc}$ & The number of subchannels.  \\
		\hline
		${{\delta }^{j}}_{COL} $ &  \multicolumn{1}{m{6.6cm}|}{The probability of a data packet collision between vehicle $i$ and vehicle $j$.} \\
		\hline
		\( b_q \) &\multicolumn{1}{m{6.6cm}|}{A vector of a two-dimensional differential equation which describes the age evolution at state \( q \). }\\
		\hline
		${P}_{O}$ &\multicolumn{1}{m{6.6cm}|}{ The probability that the selection windows of vehicle $i$ and $j$ overlapped. } \\
		\hline
		${P}_{SH|O}$ &\multicolumn{1}{m{6.6cm}|}{The probability that vehicle $i$ and $j$ choose resource from the overlapped window} \\
		\hline
		\( p_{i,j} \) &\multicolumn{1}{m{6.6cm}|}{The average rate at which link \( i \) is preempted by link \( j \).} \\
		\hline
		${N}_{Sh}$ &\multicolumn{1}{m{6.6cm}|}{The amount of shared resources within the overlapped selection window.} \\
		\hline
		${{\delta }^{j}}_{HD}$ & \multicolumn{1}{m{6.6cm}|}{The probability that both vehicles transmit data simultaneously.} \\
		\hline
	\end{tabular}
	\vspace{-0.6cm}
    \label{tab1}
\end{table}
\subsection{Transmission rate}

To ensure fairness access means that vehicles moving with varying speeds should transmit the same data quantity when covered by the RSU. This requirement can therefore be formulated as:
\begin{equation}
	 {E[Bit^{i}]=C},
	\label{eq1}
\end{equation}
where $Bit^{i}$ denotes the data size transmitted by vehicle $i$. $C$ is a constant, considering the possibility of transmission failure, we take the expected value. Specifically, $Bit^{i}$ is given by:
\begin{equation}
	{Bit^{i}={{C}_{i}}\cdot {{T}_{i}} }.
	\label{eq2}
\end{equation}
We set ${{P}_{PRR}}^{_{i}}$ as the probability of successful transmission. Thus, Eq. \eqref{eq1} is given by:
\begin{equation}
{{C}_{i}}\cdot {{T}_{i}}\cdot {{P}_{PRR}}^{_{i}}=C,
\label{eq3}
\end{equation}
where ${C}_{i}$ denotes vehicle's transmission rate, while ${T}_{i}$ represents the duration vehicle $i$ remains within the RSU's coverage area. Thus, ${T}_{i}$ is given by:
\begin{equation}
{{T}_{i}}=\frac{R}{{{v}_{i}}},
\label{eq4}
\end{equation}
where $R$ indicates RSU's range, and ${v}_{i}$ represents the vehicle's velocity. Considering the upper limit of our presented framework, following Shannon theory, ${C}_{i}$ can be mathematically formulated as:
\begin{equation}
{{C}_{i}}=B\cdot{{\log }_{2}}(1+\frac{{{p}_{i}}\cdot {{h}_{i,r}}\cdot {{({{d}_{i,r}})}^{-\partial }}}{{{\sigma }^{2}}}),
\label{eq5}
\end{equation}
where $B$ denotes system bandwidth, ${\partial }$ denotes the path loss index, ${p}_{i}$ denotes the transmission power of vehicle $i$. ${d}_{i,r}$ is the geometric distance between vehicle $i$ and RSU. ${h}_{i,r}$ is the channel gain between vehicle $i$ and RSU. ${\sigma }^{2}$ refers to the additive white Gaussian noise power. The distance ${d}_{i,r}$ can be described as:

\begin{equation}
{d}_{i,r}=\left\| {{P}_{o}}^{i}-{{P}_{o}}^{r} \right\|,
\label{eq6}
\end{equation}
where ${P}_{o}^{i}$ denotes the vehicle $i$'s position, and ${P}_{o}^{r}$ is the RSU's position.

Based on \cite{12}, we apply the Autoregressive model \cite{53} which characterizes the temporal dependency between consecutive channel gains ${h}_{i,r}$ and ${h}_{i,r}'$:
\begin{equation}
{h}_{i,r}={{\rho }_{i}}\cdot{h}_{i,r}'+e(t)\cdot\sqrt{1-\rho _{i}^{2}},
\label{eq8}	
\end{equation}
where ${\rho }_{i}$ is the autocorrelation coefficient, ${h}_{i,r}'$ is the channel gains at previous slot, while the error vector $e(t)$ follows a Gaussian distribution. In addition, to model vehicular mobility-induced Doppler spread \cite{54}, we employ Jake’s fading spectrum, ${{\rho }_{i}}={{J}_{0}}(2\pi f_{d}^{i}t)$, where ${{J}_{0}}(\cdot )$ denotes the first-kind zeroth-order Bessel function. $f_{d}^{i}t$ represents the Doppler frequency determined by:
\begin{equation}
	f_{d}^{i}=\frac{v_i}{{{\Lambda }_{0}}}\cdot\cos \theta ,
	\label{eq9}	
\end{equation}
where ${\Lambda }_{0}$ denotes the wavelength, and $\cos \theta $ indicates the cosine value between the vehicle's velocity vector and the signal propagation direction.

\subsection{Successful receiving probability}
Then, we focus on the ${{P}_{PRR}}^{_{i}}$.
${P_{PRR}}^i$ quantifies the successful receiving probability of vehicle $i$ transmissions at the RSU, mathematically defined as:

\begin{equation}
	{{P}_{PRR}}^{_{i}}=\prod\limits_{j\ne \text{i}}{(1-{{\delta }^{j}}_{COL})\cdot }\prod\limits_{j\ne \text{i}}{(1-{{\delta }^{j}}_{HD})}.
	\label{eq10}	
\end{equation}
The collision probability  ${{\delta }^{j}}_{COL} $ characterizes Physical Resource Block (PRB) allocation conflicts between vehicle $i$ and interfering vehicle $j$ where multiple transmitters may share the same PRBs, when multiple vehicles attempt to select resources at nearly the same time. Let ${{\delta }^{j}}_{HD}$ denote the probability that vehicles transmit simultaneously. Following the collision probability in \cite{13}, ${{\delta }^{j}}_{COL} $  is defined through:
\begin{equation}
	{{\delta }^{j}}_{COL}={{P}_{O}}\cdot{{P}_{SH|O}}\cdot\frac{{{C}_{Ca}}}{N_{Ca}^{2}},
	\label{eq11}	
\end{equation}
where ${P}_{O}$ denotes the probability that the selection window for vehicle $i$ and $j$   overlap. Meanwhile, ${P}_{SH|O}$ represents the chance that vehicles choose resources from this overlapped window. In cases where an overlap occurs, ${C}_{Ca}$ expresses the candidate PRBs that vehicles have in common, while $N_{Ca}$ stands for the total average number of candidate PRBs available.
The mathematical expression for ${P}_{O}$ is given by:
\begin{equation}
{{P}_{O}}=\frac{{{w}_{i}}+{{w}_{j}}+1}{1000\cdot {{2}^{\mu }}\cdot RRI},
	\label{eq12}	
\end{equation}
where ${w}_{i}$ and ${w}_{j}$ denotes the selection windows size of vehicle $i$ and $j$ respectively. $RRI$ represents the resource selection interval. $\mu$ represents the subcarrier spacing coefficient.
${P}_{SH|O}$ is given by:
\begin{equation}
	{{P}_{SH|O}}={{(\frac{{{N}_{Sc}}\cdot{{N}_{Sh}}}{{{N}_{r}}})}^{2}},
	\label{eq13}	
\end{equation}
where ${N}_{Sc}$ denotes the total number of subchannels, while ${N}_{Sh}$ refers to the number of resources that are common within the overlapping selection window. ${N}_{r}$ is the total number of resources. Specifically, ${N}_{Sh}$ can be defined as:
\begin{equation}
	{{N}_{Sh}}=\frac{({{w}_{i}}+1)({{w}_{j}}+1)}{{{w}_{i}}+{{w}_{j}}+1}.
	\label{eq14}	
\end{equation}

Owing to the half-duplex constraint in vehicular communications, if vehicles transmit simultaneously, the receiver will be unable to decode the data packet, resulting in a error transmission. According to \cite{13}, ${{\delta }^{j}}_{HD}$ is given by:
\begin{equation}
{{\delta }^{j}}_{HD}=\frac{{{\tau }_{j}}}{1000},
\label{eq15}	
\end{equation}
where $\tau_j$ indicates the rate at which vehicle $j$ generates packets. Consequently, ${P_{PRR}}^i$ is related with $w$.

\subsection{Fairness index}
According to the analysis above, Eq. \eqref{eq3} is reformulated as:

\begin{equation}
\begin{aligned} 
C=&B\cdot{{\log }_{2}}(1+\frac{{{p}_{i}}\cdot {{h}_{i,r}}\cdot {{({{d}_{i,r}})}^{-\partial }}}{{{\sigma }^{2}}})\cdot \frac{R}{{{v}_{i}}}\\ 
&\cdot \prod\limits_{j\ne \text{i}}{(1-{{\delta }^{j}}_{COL})\cdot }\prod\limits_{j\ne \text{i}}{(1-{{\delta }^{j}}_{HD})}.
\label{eq16}
\end{aligned}	
\end{equation}
By discarding terms that are not related to vehicle \( i \), Eq.\eqref{eq16} can be rewritten as: 
\begin{equation}
\begin{aligned} 
K_{index}^{i}=\frac{C}{{{C}'}}=&{{\log }_{2}}(1+\frac{{{p}_{i}}\cdot {{h}_{i,r}}\cdot {{({{d}_{i,r}})}^{-\partial }}}{{{\sigma }^{2}}})\\
&\cdot \frac{\prod\limits_{j\ne \text{i}}{(1-{{\delta }^{j}}_{COL})}}{{{v}_{i}}},
\label{eq17}
\end{aligned}	
\end{equation}
where 
\begin{equation}
{C}'=B\cdot R\cdot \prod\limits_{j\ne \text{i}}{(1-{{\delta }^{j}}_{HD})}.
\label{eq18}	
\end{equation}

As a result, we now formulate a fairness metric for vehicle $i$. In addition, due to the fact that  \({{d}_{i,r}}\) is related with \(v_i\) while \({{P}_{PRR}}^{i}\) depends on the selection window \(w\), the fairness index \(K_{index}^{i}\) is influenced by \(v\) and \(w\). This means that if we get a vehicle's velocity, the \(w\) can be adaptively adjusted to improve fairness.

Furthermore, through computing the average value of $K_{index}^{i}$ for all the vehicles within the RSU's range, we can get:

\begin{equation}
K_{index}=\frac{\sum\limits_{i=1}^{N}{{{{K_{index}^{i}}}}}}{N}.
\label{eq19}	
\end{equation}
The overall average fairness of the network is quantified by the index $K_{index}$. When an individual vehicle’s fairness index \(K_{index}^{i}\) closely aligns with \(K_{index}\), it signifies that the vehicle is accessing communication in a fair manner.

So far, we have only achieved fair access among vehicles by adjusting the selection window size. However, adjusting the selection window size may lead to an increase in the AoI of vehicles. In particular, considering the existence of the preemption mechanism, when high-priority vehicles frequently preempt the communication resources of low-priority vehicles, the AoI of the preempted vehicles may significantly increase due to waiting for resources. At the same time, in order to achieve fairness, it may be necessary to enlarge the selection window size for certain vehicles, which can likewise cause their AoI to grow. Therefore, in the following, we will model and analyze the AoI under the preemption mechanism.

\section{Age of Information}
\label{sec5}
In this section, we further analyze the average AoI in the proposed model. The AoI refers to the mean age of the data exchanged between each vehicle and the RSU. The communication relationships among vehicles and RSU are referred as transmission links, where link \( k \) denotes the communication link between vehicle \( k \) and the RSU. 

We employ the SHS to design the system transition\cite{22,23}. The data sampling time is assumed to be negligible, as it is significantly smaller compared to the data transmission time \cite{17}.

Next, we model the transmission process. We define the state set as \( (q(t), \boldsymbol{x(t))} \), where \( q(t) \in \{0, 1, 2, \dots, N\} \) represents the system state at time slot \( t \), and \( N \) denotes the amount of vehicles. Specifically, \( q(t) = 0 \) stands for that no transmission occurs at time \( t \), i.e. the channel is idle, whereas \( q(t) = k \) indicates that link \( k \) captures the channel and is transmitting data. 

We set \( \boldsymbol{x(t)} = [x_0(t), x_1(t)] \) as the AoI of link \( k \), where \( x_0(t) \) denotes the AoI at the RSU, and \( x_1(t) \) denotes the AoI of data generated and transmitted at vehicle \( k \) at time \( t \). 

At system initialization, the AoI at the RSU \( x_0(t) \) is set to 0 and then begins increasing  with unit slope. When a data packet from link \( k \) is received, \( x_0(t) \) is reset to the AoI of link \( k \). Similarly, when vehicle \( k \) generates a data packet, the AoI at the sender \( x_1(t) \) is initialized to 0. Next, it begins increasing with unit slope until the data is successfully received by the RSU. 

Thus, based on the above definitions, \( q(t) \) can be mapped as a discrete process, while \( \boldsymbol{x(t)} \) can be mapped as a continuous process.

\begin{table}[tbp]
	\centering
	\resizebox{\columnwidth}{!}{%
		\begin{tabular}{cccccc}
			\hline
			\boldsymbol{$l$} & \boldsymbol{$q_l \rightarrow q_l'$} & \boldsymbol{$\lambda^{(l)}$} & \boldsymbol{$x' = xA_l$} & \boldsymbol{$A_l$} & \boldsymbol{$\overline{v}_{q_l'} = \overline{v}_{q_l} A_l$} \\
			\hline
			$1$ & $0 \rightarrow 1$ & $R_1$ & $[x_0, x_1]$ & $\begin{bmatrix}1 & 0 \\ 0 & 1\end{bmatrix}$ & $[\overline{v}_{00}, \overline{v}_{01}]$ \\
			$\vdots$ & $\vdots$ & $\vdots$ & $\vdots$ & $\vdots$ & $\vdots$ \\
			$N$ & $0 \rightarrow N$ & $R_{N}$ & $[x_0, x_1]$ & $\begin{bmatrix}1 & 0 \\ 0 & 1\end{bmatrix}$ & $[\overline{v}_{00}, \overline{v}_{01}]$ \\
			\hline
			$N + 1$ & $1 \rightarrow 0$ & $H_1$ & $[x_0, x_1]$ & $\begin{bmatrix}0 & 1 \\ 1 & 0\end{bmatrix}$ & $[\overline{v}_{10}, \overline{v}_{11}]$ \\
			$\vdots$ & $\vdots$ & $\vdots$ & $\vdots$ & $\vdots$ & $\vdots$ \\
			$N + k$ & $k \rightarrow 0$ & $H_k$ & $[x_1, 0]$ & $\begin{bmatrix}0 & 1 \\ 0 & 0\end{bmatrix}$ & $[\overline{v}_{k1}, 0]$ \\
			$\vdots$ & $\vdots$ & $\vdots$ & $\vdots$ & $\vdots$ & $\vdots$ \\
			$2N$ & $N \rightarrow 0$ & $H_{N}$ & $[x_0, x_1]$ & $\begin{bmatrix}1 & 0 \\ 1 & 1\end{bmatrix}$ & $[\overline{v}_{N0}, \overline{v}_{N1}]$ \\
			\hline
			\multicolumn{1}{c}{$2N+1$} & $1 \rightarrow 2$ & $P_{1,2}$ & $[x_0, x_1]$ & $\begin{bmatrix}1 & 0 \\ 0 & 1\end{bmatrix}$ & $[\overline{v}_{10}, \overline{v}_{11}]$ \\
			
			$\vdots$ & $\vdots$ & $\vdots$ & $\vdots$ & $\vdots$ & $\vdots$ \\
			$3N-1$& $1 \rightarrow N$ & $P_{1,N}$ & $[x_0, x_1]$ & $\begin{bmatrix}1 & 0 \\ 0 & 1\end{bmatrix}$ & $[\overline{v}_{10}, \overline{v}_{11}]$ \\
			\hline
			$(k+1)N-k+2$ & $k \rightarrow 1$ & $P_{k,1}$ & $[x_0, 0]$ & $\begin{bmatrix}1 & 0 \\ 0 & 0\end{bmatrix}$ & $[\overline{v}_{k0}, 0]$ \\
		
			$\vdots$ & $\vdots$ & $\vdots$ & $\vdots$ & $\vdots$ & $\vdots$ \\
			$(k+2)N-k$& $k \rightarrow N$ & $P_{k,N}$ & $[x_0, 0]$ & $\begin{bmatrix}1 & 0 \\ 0 & 0\end{bmatrix}$ & $[\overline{v}_{k0}, 0]$ \\
			\hline
			$N^2+2$ & $N \rightarrow 1$ & $P_{N,1}$ & $[x_0, x_1]$ & $\begin{bmatrix}1 & 0 \\ 0 & 1\end{bmatrix}$ & $[\overline{v}_{N0}, \overline{v}_{N1}]$ \\
			
			$\vdots$ & $\vdots$ & $\vdots$ & $\vdots$ & $\vdots$ & $\vdots$ \\
			$N^2+N$& $N \rightarrow N - 1$ & $P_{N,N-1}$ & $[x_0, x_1]$ & $\begin{bmatrix}1 & 0 \\ 0 & 1\end{bmatrix}$ & $[\overline{v}_{N0}, \overline{v}_{N1}]$ \\
			\hline
		\end{tabular}%
	}
	\caption{SHS Transitions model }
	\label{tab2}
\end{table}

\begin{figure}[tbp]
	\centering
	\includegraphics[width=\linewidth, scale=1.00]{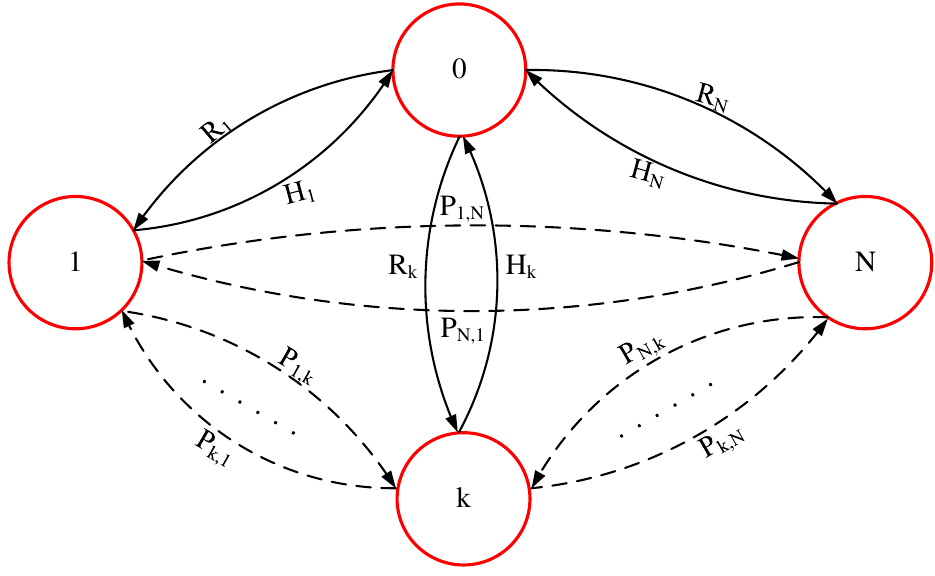}
	\caption{Markov Model}
	\label{fig1}
	\vspace{-0.0cm}
\end{figure}

In a Markov chain, transitions occur between multiple states. According to the SHS framework, transitions between discrete states \( q(t) \) will trigger resets of the continuous process \( \boldsymbol{x(t)} \). This reset process is given by \( \boldsymbol{x' = x A_l} \), where \( l \) represents the transition of the discrete process, and \(\boldsymbol{ A_L} \) represents the reset mapping for that transition. Here, \( \boldsymbol{x} \) and \( \boldsymbol{x'} \) denote the continuous process values before and after the reset, while \( q_l \) and \( q_l' \) represent the discrete states before and after the transition. 

We set \( \lambda \) as the transition rate of the discrete process \( l \). Additionally, \( \boldsymbol{v_{q_l}} = [v_{q_0}, v_{q_1}] \) represents the stationary correlation between \( q_l \) and \( \boldsymbol{x} = [x_0, x_1] \). Specifically, \( v_{q_0} \) denotes the stationary correlation between \( q_l \) and \( x_0 \), while \( v_{q_1} \) represents the correlation between \( q_l \) and \( x_1 \). Given \( \boldsymbol{x' = x A_l} \), we can derive \( \boldsymbol{v_{q_l'} = v_{q_l} A_l} \), where \( \boldsymbol{v_{q_l'} }\) represents the stationary correlation between \( q_l' \) and the reset process \( \boldsymbol{x'} \).

Furthermore, we define the average service rate of link \( k \) as \( H_k \) while the average failed transmission rate as \( R_k \). Considering the preemption mechanism specific to the SPS in NR V2X, we introduce \( p_{i,j} \) to represent the average rate at which link \( i \) is preempted by link \( j \). When link \( k \) successfully transmits, the channel transitions to an idle state, and the original state \( k \) transitions to state \( 0 \) at a rate of \( H_k \). Conversely, if link \( k \) fails to transmit, due to the retransmission mechanism in SPS, link \( k \) will recapture the channel, causing state \( 0 \) to transition back to state \( k \) at a rate of \( R_k \). Additionally, when link \( i \) is preempted by link \( j \), state \( i \) transitions to state \( j \) at a rate of \( p_{i,j} \).

According to Fig. 3, we can summarize the SHS transitions in Table II.

We now provide a detailed explanation of the transitions listed in Table II:
\begin{itemize}
	
\item[1)] Transition \( l_1 \) (\( l_1 =  \{1, 2, 3, \dots, N\} \)), representing the scenario where an idle channel is occupied, and a transition occurs on link \( k \) at a rate of \( R_k \). This implies that the previous transmission has failed. Thus, the AoI remains unchanged. As a result, we have:  
\[
\boldsymbol{x' = x A_{l_1}} = [x_0, x_1], \quad \boldsymbol{v_{q_{l_1}'} = v_{q_{l_1}} A_{l_1}} = [v_{00}, v_{01}].
\]

\item[2)] Transition \( l_2 \)  (\( l_2 =  \{N+1, N+2, \dots, 2N\} \)), indicating that the channel enters an idle state and transitions on link \( N+k \) at a rate of \( H_k \). This corresponds to a successful transmission, where the \( x_1 \) is reset to \( x_1 \), while the AoI of vehicle $k$ is reset to \( 0 \). Thus, we obtain:  
\[
\boldsymbol{x' = x A_{l_2}} = [x_1, 0], \quad \boldsymbol{v_{q_{l_2}'} = v_{q_{l_2}} A_{l_2}} = [v_{k1}, 0].
\]  
For transitions other than \( N+k \), a successful transmission on any other link does not reset the AoI of link \( k \). Hence, we have:  
\[
\boldsymbol{x' = x A_{l_2} = [x_0, x_1]}, \quad \boldsymbol{v_{q_{l_2}'} = v_{q_{l_2}} A_{l_2}}.
\]

\item[3)] Transition \( l_3 \)  (\( l_3 =  \{2N+1, 2N+2, \dots, N^2 + N\} \)), representing the preemption process occurring at a rate of \( p_{i,j} \). During this transition, the RSU's AoI does not reset. The AoI of the link is reset to \( 0 \) and begins linear growth with a slope of \( 1 \) because when the vehicle being preempted estimates that its resources will be used, it will release the resources, that is, when the preempt vehicle's packets are generated. Thus, we have:  
\[
\boldsymbol{x' = x A_{l_3}} = [x_0, 0], \quad \boldsymbol{v_{q_{l_3}'} = v_{q_{l_3}} A_{l_3}} = [v_{k0}, 0].
\]  
Similarly, for transitions other than \( n+k \), a successful transmission on any other link does not reset the AoI of link \( k \). Therefore, we obtain:  
\[
\boldsymbol{x' = x A_{l_3}} = [x_0, x_1], \quad \boldsymbol{v_{q_{l_3}'} = v_{q_{l_3}} A_{l_3}}.
\]
\end{itemize}

According to \cite{23}, the average AoI for link \( k \) is calculated as:
\begin{equation} 
	\bar{\Delta}_k = \sum\bar{v}_{q0}, \quad \forall k \in {1,2,\dots,N}, 
	\label{20}
\end{equation}
According to this formula, to compute the average AoI \( \bar{\Delta}_k \) for link \( k \), it is essential to derive \( v_{q0} \).

Firstly, based on the analytical approach provided in \cite{23},
\begin{equation}
	\begin{aligned}
		 \boldsymbol{{{{\bar{v}}}_{q{{l}_{a}}}}} \left( \sum\limits_{{{l}_{a}}}{{{\lambda }^{({{l}_{a}})}}} \right) 
		= \boldsymbol{{{b}_{q}}} \bar{\pi }_q + \sum\limits_{{{l}_{b}}}{{{\lambda }^{({{l}_{b}})}}}\boldsymbol{{{{\bar{v}}}_{q{{l}_{b}}}}{{A}_{{{l}_{b}}}}},   
		&  {{l}_{a}} \in {{L}_{q}}, \\
		&  {{l}_{b}} \in {{{{L}'}}_{q}},
	\end{aligned}
	\label{21}
\end{equation}
where \(\boldsymbol{ b_q} \) denotes a vector of a two-dimensional differential equation describing the age evolution in state \( q \), and \( \pi_q \) denotes the stationary probability  of state \( q \). Additionally, \( l_a \) and \( l_b \) refer to the discrete states sets before and after the transition.

Since vehicle \( k \) generates data packets and initiates transmissions only upon capturing the channel, there are no data packets for link \( k \) in the system network unless the state equals \( k \). Consequently, the AoI on link \( k \) increases linearly at a unit rate when \( q = k \) and remains at zero otherwise. Simultaneously, the AoI at the RSU always grows linearly at a unit rate. Based on this analysis, we derive the following:

\begin{equation}
\boldsymbol{b_q} =
\begin{cases} 
	[1,0], & \text{for } \forall q \neq k, \\ 
	[1,1], & \text{for } \forall q = k.
\end{cases}
	\label{22}
\end{equation}
Then, in order to apply Eq. \eqref{21}, we need to get the stationary distribution of state \(q\). Based on \cite{23}, the stationary distribution can be obtained as follows: 
\begin{equation}
	\begin{aligned}
		{{{\bar{\pi }}}_{\bar{q}}} \sum\limits_{l\in {{L}_{q}}}{{{\lambda }^{(l)}}} &= \sum\limits_{l\in {{{{L}'}}_{q}}}{{{\lambda }^{(l)}}}{{{\bar{\pi }}}_{q_l}}, \quad \bar{q}\in Q, \\ 
		\sum\limits_{\bar{q}\in Q}{{{{\bar{\pi }}}_{\bar{q}}}} &= 1.
	\end{aligned}
	\label{23}
\end{equation}
By solving the above equations, we can derive:
\begin{equation}
	\begin{cases} 
		{{{\bar{\pi }}}_{0}} = \frac{1}{C(R)}, \\ 
		{{{\bar{\pi }}}_{k}} = \frac{{{R}_{k}}}{C(R)({{H}_{k}}-\sum\limits_{j \neq k}{{{p}_{j,k}}})},
	\end{cases}
	\label{24}
\end{equation}
where $C(R)$ is a normalization factor:
\begin{equation}
C(R)=1+\sum\limits_{k=1}^{N}{\frac{{{R}_{k}}}{{{H}_{k}}-\sum\limits_{j\ne k}{{{p}_{j,k}}}}}.
	\label{25}
\end{equation}

Our goal is to apply Eq. \eqref{21} to obtain \( v_{q0} \). 
When \( q = 0 \), based on Eq.\eqref{22}, we know that \( b_q = [1,0] \). In this case, according to \cite{9}, the left side of Eq. \eqref{21} represents the transition from other states to states except \( q = 0 \), that is, the transitions from  1  to $N$ and \( 2N + 1 \) to \( N^2 + N \) as shown in Table II. The right part represents the transition from other states to state 0, that is, the transitions from \( N + 1 \) to \( 2N \) as shown in Table II. 

Furthermore, according to \cite{23}, we know that Eq. \eqref{21} applies to any set of reset mappings \( \{\boldsymbol{A_l}\} \). Therefore, Eq. \eqref{21} is applicable to the transitions from \( 2N + 1 \) to \( N^2 + N \) caused by the preemption mechanism. 

In summary, combining with Table II, we can derive:

\begin{equation}
	\begin{aligned}
{{\bar{v}}_{00}}(\sum\limits_{j=1}^{n}{{{R}_{j}}})+\sum\limits_{\begin{smallmatrix} 
		j=1 \\ 
		j\ne k 
\end{smallmatrix}}^{N}{{{{\bar{v}}}_{j0}}}\cdot \sum\limits_{i\ne j}^{N-1}{{{p}_{i,j}}+}{{\bar{v}}_{k0}}\sum\limits_{i\ne j}{{{p}_{k,i}}}\\
={{\bar{\pi }}_{0}}+\sum\limits_{\begin{smallmatrix} 
		j=1 \\ 
		j\ne k 
\end{smallmatrix}}^{N}{{{H}_{j}}}{{\bar{v}}_{j0}}+{{H}_{k}}{{\bar{v}}_{k1}},
	\end{aligned}
	\label{26}
\end{equation}

\begin{equation}
{{\bar{v}}_{01}}(\sum\limits_{j=1}^{N}{{{R}_{j}}})+\sum\limits_{\begin{smallmatrix} 
		j=1 \\ 
		j\ne k 
\end{smallmatrix}}^{N}{{{{\bar{v}}}_{j1}}}\cdot \sum\limits_{i\ne j}^{N-1}{{{p}_{j,i}}}=\sum\limits_{\begin{smallmatrix} 
		j=1 \\ 
		j\ne k 
\end{smallmatrix}}^{N}{{{H}_{j}}}{{\bar{v}}_{j1}}.
	\label{27}
\end{equation}

Moreover, the left part represents transitions from other states to state 0 when \(q\neq 0\), that are transitions \(N+1\) to \(2N\) as shown. The right part represents transitions from other states to state \(k\), i.e. \(1\) to \(N\) and from \(2N+1\) to \(N^2+N\) as shown in Table II. Therefore, based on Table II, we can obtain:
\begin{equation}
{{\bar{v}}_{q0}}\cdot {{H}_{q}}={{\bar{\pi }}_{q}}+{{R}_{q}}\cdot {{\bar{v}}_{00}}+\sum\limits_{j\ne q}^{N-1}{{{p}_{q,j}}}{{\bar{v}}_{q0}},for\text{ }\forall q=\{1,2,...,N\}
	\label{28}
\end{equation}

\begin{equation}
{{\bar{v}}_{q1}}\cdot {{H}_{q}}={{R}_{q}}\cdot {{\bar{v}}_{01}}+{{\bar{v}}_{q1}}\sum\limits_{j\ne q}^{N-1}{{{p}_{q,j}}},for\text{ }q\ne k
	\label{29}
\end{equation}

\begin{equation}
{{\bar{v}}_{k1}}\cdot {{H}_{k}}={{\bar{\pi }}_{k}}+{{R}_{k}}\cdot {{\bar{v}}_{01}}+{{\bar{v}}_{k1}}\sum\limits_{j\ne k}^{N-1}{{{p}_{k,j}}},for\text{ }q=k
	\label{30}
\end{equation}

Based on the formula derived above, we will next derive \( v_{00} \) and \( v_{q0} \). According to Eq. \eqref{28}, we can obtain ${\bar{v}}_{q0}$ when \( q \neq 0 \):

\begin{equation}
{{\bar{v}}_{q0}}=\frac{{{{\bar{\pi }}}_{q}}+{{R}_{q}}\cdot {{{\bar{v}}}_{00}}}{{{H}_{q}}-\sum\limits_{j\ne q}^{N-1}{{{p}_{q,j}}}}.
	\label{31}
\end{equation}	
According to Eq. \eqref{20}, to determine the AoI in the network, we still need to obtain \( v_{00} \). From Eq. \eqref{26}, obtaining \( v_{00} \) requires determining \( v_{k1} \). Based on Eq. \eqref{29} and Eq. \eqref{30}, we can obtain:
\begin{equation}
{{\bar{v}}_{q1}}=\frac{{{R}_{q}}\cdot {{{\bar{v}}}_{01}}}{{{H}_{q}}-\sum\limits_{j\ne q}^{N-1}{{{p}_{q,j}}}},
	\label{32}
\end{equation}
\begin{equation}
{{\bar{v}}_{k1}}=\frac{{{{\bar{\pi }}}_{k}}+{{R}_{k}}\cdot {{{\bar{v}}}_{01}}}{{{H}_{k}}-\sum\limits_{j\ne k}^{N-1}{{{p}_{k,j}}}}.
	\label{33}
\end{equation}

By combining Eq. \eqref{32} with Eq. \eqref{27}, we obtain:
\begin{equation} 
{{\bar{v}}_{01}}\sum\limits_{j=1}^{N}{{{R}_{j}}={{{\bar{v}}}_{01}}\cdot }\sum\limits_{\begin{smallmatrix} 
		q=1 \\ 
		q\ne k 
\end{smallmatrix}}^{N}{{{R}_{q}}}.
	\label{34}
\end{equation}
Therefore, ${{\bar{v}}_{01}}=0$. Combining Eq. \eqref{34} with Eq. \eqref{33}, we can obtain:
\begin{equation}
{{\bar{v}}_{k1}}=\frac{{{{\bar{\pi }}}_{k}}}{{{H}_{k}}-\sum\limits_{j\ne k}^{N-1}{{{p}_{k,j}}}}.
	\label{35}
\end{equation}
Finally, by combining Eq. \eqref{35} and Eq. \eqref{31} with Eq. \eqref{26}, we can obtain:
\begin{equation}
{{\bar{v}}_{00}}=\frac{{{H}_{k}}-\sum\limits_{j\ne k}^{N-1}{{{p}_{k,j}}}}{{{H}_{k}}\cdot {{R}_{k}}}.	
	\label{36}
\end{equation}

At this point, we have derived \( v_{q0} \). Substituting it into Eq. \eqref{20}, we derive the AoI for link \( k \):
\begin{equation}
	\begin{aligned}
		\bar{\Delta }_{k} &= \sum_{q=0}^{N} \bar{v}_{q0}, \forall k \in \{1, 2, \dots, N\} \\
		&= \bar{v}_{00} + \sum_{q=1}^{N} \bar{v}_{q0} \\
		&= \bar{v}_{00} \left[ \sum_{q=1}^{N} \frac{\bar{\pi}_{q} + R_{q} \cdot \bar{v}_{00}}{H_{q} - \sum_{j \ne q}^{N-1} p_{q,j}} \right] \\
		&= \frac{H_k - \sum_{j \ne k}^{N-1} p_{k,j}}{H_k \cdot R_k} \left[ 1 + \sum_{q=1}^{N} \frac{R_q}{H_q - \sum_{j \ne q}^{N-1} p_{q,j}} \right] \\
		&\quad + \sum_{q=1}^{N} \frac{\bar{\pi}_q}{H_q - \sum_{j \ne q}^{N-1} p_{q,j}}   \\
		&= \frac{H_k - \sum_{j \ne k}^{N-1} p_{k,j}}{H_k \cdot R_k}\cdot C(R)+ \sum_{q=1}^{N} \frac{\bar{\pi}_q}{H_q - \sum_{j \ne q}^{N-1} p_{q,j}}.  
	\end{aligned}
	\label{37}
\end{equation}

Next, by summing the AoI for all links in the network and taking the average, we get the average AoI as:
\begin{equation}
\bar{\Delta }=\frac{\sum\limits_{k=1}^{N}{{{{\bar{\Delta }}}_{k}}}}{N}.
	\label{38}
\end{equation}

According to Eq. \eqref{37}, the AoI in the network is determined by the transition rates of successful and failed vehicle transmissions, as well as the preemption process, i.e. \( H_i \), \( R_i \), \( p_{i,j} \). Therefore, the next step is to further derive \( H_i \), \( R_i \), and \( p_{i,j} \).

\textit{1) Average Service Rate:} Based on\cite{18}, the average service rate is given by:
\begin{equation}
{{H}_{i}}=\frac{1}{{{T}_{s}}},for\text{ }\forall i\in \{1,2,....N\},
	\label{39}
\end{equation}	
where \( T_s \) denotes the average successful transmission time. According to \cite{19}, it can be expressed as:
\begin{equation}
{{T}_{s}^i}={{T}_{ini}^i}+n\cdot {{T}_{r}^i},
	\label{40}
\end{equation}
where \( T_{\text{ini}}^i \) represents the total transmission time, and \( t_r \) denotes the time required for retransmission. Due to successful transmission, we consider the information is transmitted only once, so the retransmission time is zero. \( T_{ini}^i \) is given by:
\begin{equation}
{{T}_{ini}^i}={{t}_{sch}^i}+{{t}_{pkt}^i},
	\label{41}
\end{equation}
where \( t_{sch}^i \) represents the time required for resource scheduling, while \( t_{pkt}^i \) denotes the actual transmission time. According to Fig. 4, \( t_{sch}^i \) can be expressed as:
\begin{equation}
{{t}_{sch}^i}={{t}_{p}^i}+{{t}_{fa}^i}+{{t}_{w}^i},
	\label{41}
\end{equation}
\begin{figure}[tbp]
	\centering
	\includegraphics[width=\linewidth, scale=1.00]{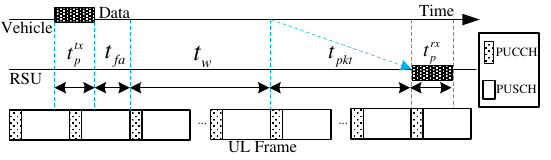}
	\caption{Lantency Model}
	\label{fig1}
	\vspace{-0.0cm}
\end{figure}
where \( {t}_{p}^i \) represents the time required for the sender to process the data, which, according to [22], depends on the vehicle's computational processing capability. Within the Uplink (UL), the Physical UL Control Channel (PUCCH) transmission always occurs in a slot's last symbol, with the remaining symbols allocated for the Physical Uplink Shared Channel (PUSCH) to transmit. Therefore, \( t_{fa}^i \) is constrained by the time slot, which based on digital numerology, ranges from 1 ms to 0.0625 ms. \( t_w \) refers to the time required for resource scheduling, i.e. the size of the selection window. \( t_{\text{pkt}}^i \) is the time actually used for data transmission can be expressed as:
\begin{equation}
{{t}_{pkt}^i}=\frac{Bit}{C_i},
	\label{43}
\end{equation}
where $Bit$ denotes the packet size.

\textit{2) Average Failure Rate:} Similarly, the average transition rate when link \( i \) fails to transmit is given by:
\begin{equation}
{{R}_{i}}=\frac{1}{{{T}_{f}^i}},for\text{ }\forall i\in \{1,2,....N\},	
	\label{44}
\end{equation}
where \( T_f \) denotes the average transmission failure time:
\begin{equation}
{{T}_{f}^i}={{T}_{ini}^i}+n\cdot {{T}_{r}^i},
	\label{45}
\end{equation}
where \( T_{\text{ini}}^i \) is the same as for a successful transmission. Since the SPS scheduling mechanism uses the HARQ retransmission mechanism, and due to the inability to predict whether or when the RSU may need a retransmission, dynamic scheduling is employed. If the packet transmit fail, a retransmission occurs, and the RSU transmit a negative acknowledgment (NACK). Because of the additional transmission, a delay $t_{\text{NACK}}$ is introduced. Therefore, \( {T}_{r}^i\) can be expressed as:
\begin{equation}
{{T}_{r}^i}={{t}_{NACK}^i}+{{t}_{sch}^i}+{{t}_{pkt}^i},
	\label{46}
\end{equation}

\begin{equation}
{{t}_{NACK}^i}={{t}_{p}^i}+{{t}_{fa}^i}+{{t}_{pkt}^i}.
	\label{47}
\end{equation}

\textit{3) Average preemption Rate:}
Next, we consider the average transition rate during the preemption process. 

According to \cite{1}, the preemption process occurs when the preempting side generates a data packet. Once the preempting side identifies its higher traffic priority, the preempted side releases the resources for the preempting side to choose. Therefore, the average transition time for the preemption process can be expressed as:
\begin{equation}
{{T}_{p}}_{_{i,j}}={{T}_{sch}^i}+{{t}_{p}^j},
	\label{48}
\end{equation}
where ${T}_{sch}^i$ denote the time required for resource scheduling of link \( i \). Therefore, the average transition rate of the preemption process can be expressed as:
\begin{equation}
{{P}_{i,j}}=\frac{1}{{{T}_{{{p}_{i,j}}}}}.
	\label{49}
\end{equation}

\section{Optimization Problem and Solution}
\label{sec6}
This section, we presents the formulation of a joint optimization problem for fair access and AoI based on section \ref{sec4} and \ref{sec5}. To solve this problem, we propose an enhanced MOEA/D algorithm integrated with LLMs \cite{51}. The objective is to determine the optimal selection window size for each vehicle, thereby achieving equitable channel access across the vehicular network while jointly minimizing the network average AoI.
\subsection{Optimization Objective}
The optimization framework simultaneously addresses two critical objectives: 
\begin{itemize}
	
	\item[(1)] Fair access among vehicles.
    \item[(2)] Minimization of the network's AoI.
\end{itemize}

The decision variables are the selection window sizes of individual vehicles. A fairness index $K_{index}^i$ is defined such that when $K_{index}^i$ approximates the network’s average fairness index $K_{index}$, equal channel access is considered achieved. The mathematical formulation of this optimization problem is expressed as:


\textbf{Objectives 1 to $\boldsymbol{N}$:} To reduce the difference between each vehicle's fair index and the averaged index.

\begin{equation}
	F_{K_i}(\boldsymbol{w}) =  \left|  K_{index}(\boldsymbol{w}) - K_{index}^i(\boldsymbol{w}) \right|,i \in [1, \ldots, N],
	\label{50}
\end{equation}
$\boldsymbol{w}=\{w^1,w^2,...,w^N\}$. 

\textbf{Objective N+1:} To minimize the averaged AoI in the network. \\
\begin{equation}
	F_{age} = \min \overline{\Delta}.
	\label{51}
\end{equation}

Thus, the joint multi-objective optimization problem is given by:
\begin{equation}
	\begin{aligned}
		\min_{\boldsymbol{w}} \; &\boldsymbol{F}(\boldsymbol{w}) = [ F_{K_{1}}(\boldsymbol{w}), F_{K_{2}}(\boldsymbol {w}),\dots,F_{K_{N}}(\boldsymbol{w}), F_{age}(\boldsymbol{w})]^T
		\\
		\qquad \qquad & s.t \\
		& \boldsymbol{w}=\{w^1,w^2,...,w^N\},\\
		&w^{LB} \leq w^i  \leq w^{UB}, i \in [1, \ldots, N],
		\label{52}
	\end{aligned}
\end{equation}
where $ w^{LB}$ and $w^{UB}$ represent the lower and upper limit of the selection window sizes, according to the 3GPP standard \cite{55}. 

To solve Eq. \eqref{52}, we can get a Pareto optimal solution set, and in order to have an exact window size for each vehicle, we need to filter out an optimal solution. We formulate the filtering rules as follows: under the condition that all $F_{K_i}(\boldsymbol{w}) $ are within the bounds, the group of solutions with the smallest AoI is selected. Therefore, we can define the optimization goal as:
\begin{equation}
	\begin{aligned}
		\min_{\boldsymbol{w}} \; &F_{age}(\boldsymbol{w})		\\
		 s.t    \quad  & F_{K_i}(\boldsymbol{w}) \leq K_{bound},	i \in [1, 2, \ldots, N], \\			
		&\boldsymbol{w} \in \mathcal{P}.
		\label{53}
	\end{aligned}
\end{equation}
where $\mathcal{P}$ is the Pareto optimal solution set which is solved by Eq. \eqref{52}.To adaptively determine $K_{\text{bound}}$, we first sort all fairness deviations in ascending order, and then select the minimal deviation among the largest 10\% of them. Thus, $K_{\text{bound}}$ can be described as:

\begin{equation}
K_{bound} = \min \left\{ F_{K}^{(j)} \;\middle|\; j = \left\lceil 0.9 \mathcal{P} \right\rceil, \ldots, \mathcal{P} \right\},
\end{equation}
where $F_K^{(j)}$ represents the $j$-th smallest value in the ascendingly ordered sequence of all fairness deviations $F_{K_i}$.

Then, after solving the optimization objective, we can adjust the window size of the vehicle adaptively according to the speed of the vehicle, so as to minimize the AoI of the network under the condition of ensuring that all vehicles are close to fair access.

\subsection{Optimization Solution}
In this section, we employ a MOEA/D algorithm based on a LLM to solve the optimization problem defined in Eq. \eqref{52} \cite{51}. The algorithm inputs consist of the number of objectives, maximum iteration count, reference direction partitioning number, vehicle speed, and neighborhood size. The detailed workflow is presented in Algorithm \ref{alg1}.

\begin{algorithm}[ht]
	\caption{LLM-Based MOEA/D Algorithm }
	\label{alg1}
    \KwInput{objectives $N+1$, generations $G_{\max}$, Partition Number $n_p$, speed $v$, neighbor size $K$}
	\KwOutput{$\boldsymbol{w}^{*}$}
	
	\textbf{Initialization Phase:} \\

	\For{$i \gets 1$ \KwTo $H$}{
		$\mathbf{w}_i \gets \left(\frac{k_1}{n_p}, \frac{k_2}{n_p}, \ldots, \frac{k_H}{n_p}\right)$ \\
		\textbf{where} $k_1 + k_2 + \cdots + k_H = n_p\;$  
	}
	$\mathbf{W} \gets \{\mathbf{w}_1, \ldots, \mathbf{w}_H\}$ \;  
	
	\ForEach{$\mathbf{w}_i \in \mathbf{W}$}{
		$\mathcal{N}_i \gets \arg\underset{j \in [H]}{\mathrm{Top_K}} \cos(\mathbf{w}_i, \mathbf{w}_j)$ \;  
	}	
	$\mathcal{P} \gets \{\boldsymbol{w_1}, \ldots, \boldsymbol{w_H}\}$ \;  
	$\mathbf{z}^* \gets   \left( \min F_{K1}(\mathcal{P}), \ldots, \min F_{K_{N+1}}(\mathcal{P})\right)$ \;
	\textbf{Main Optimization Loop:}\\
	\For{$g = 1$ \KwTo $G_{\max}$}{
		\ForEach{subproblem $i \in [H]$}{

			$\mathcal{P}_{\text{parents}} \gets 
			\begin{cases}
				 \text{Select from } \mathcal{N}_i \ \text{with }  p_{\text{nei}} \\
				\text{Random selection } \text{with } 1-p_{\text{nei}}
			\end{cases}
			$
			
			\textbf{LLM-guided Crossover:} \\
			offspring $\mathbf{o} \gets \text{LLM-Mate}(\mathcal{P}_{\text{parents}})$ \;
			
			$\mathbf{z}^* \gets \min(\mathbf{z}^*, \boldsymbol{F(\mathbf{o})})$ \;
		
			\ForEach{$j \in \mathcal{N}_i$}{
				\If{$g(\mathbf{o}|\mathbf{w}_j,\mathbf{z}^*) < g(\boldsymbol{w_j}|\mathbf{w}_j,\mathbf{z}^*)$}{
					$\boldsymbol{w_j} \gets \mathbf{o}$ \;
				}
			}
		}
	}
	  initialize $\overline{\Delta} = +\infty$\;
	  \ForEach{$p \in \mathcal{P}$ }
	{
		\If{for each $p \in [1,...,N]$, $F_{K_i} \leq K_{bound}$}
		{
			\If{$F_{age} \leq \overline{\Delta}$}
			{
				$\overline{\Delta} = F_{age}$, $\boldsymbol{w}^{*} = \boldsymbol{w}^p$\;
			}
		}
	}
	return 	$\boldsymbol{w}^{*}$\;
	\SetKwProg{Proc}{Procedure}{}{}
	\Proc{$\text{LLM Crossover}(\mathcal{P}_{\text{parents}})$}{

		$\tilde{\boldsymbol{w}} \gets \frac{w-w^LB}{w^UB-w^LB},\; \forall \boldsymbol{w} \in \mathcal{P}_{\text{parents}}$ \;

		$\mathcal{T} \gets \textbf{Prompt Construction:}\left(\tilde{\boldsymbol{w}}, f(\boldsymbol{w}) \right)$ \;
		\Repeat{3 times}{
			$\mathbf{o}_{norm} \gets \text{LLM with} (\mathcal{T})$ \;
			\If{$\mathrm{Validate}(\mathbf{o}_{\text{norm}})$}{
				\textbf{break} \;
			}
		}
		
		$\mathbf{o} \gets \mathbf{o}_{norm} \cdot (w^U - w^L) + w^L$ \;
		
		\Return $\mathbf{o}$ \;
	}

\end{algorithm}


First, weight vectors $\mathbf{W} = \{\mathbf{w}_1, \ldots, \mathbf{w}_H\}$ are generated by the Das-Dennis uniform sampling scheme to decompose the multi-objective problem into $H$ subproblems, each corresponding to an optimization direction.
Cosine similarity between weight vectors is computed to select $K$ nearest neighbors as:
\begin{equation}
	\cos(\mathbf{w_i}, \mathbf{w_j}) = \frac{\mathbf{w_{i}} \cdot \mathbf{w_{j}}}{\|\mathbf{w_{i}}\| \|\mathbf{w_{j}}\|}.
	\label{53}
\end{equation}

Population initialization is performed by assigning randomly generated initial solutions to each weight vector.
The ideal point is initialized to record current optimal values of each objective function, guiding subsequent optimization directions.
(This completes Steps 1-9 of the algorithm)

In each iteration cycle:
For each subproblem, parent solutions are selected with probability $p_{nei}$ based on neighborhood relationships; otherwise, random selection is performed.
Next, we will perform LLM-guided crossover operations. Here, the LLM serves as a black-box operator used to generate a new set of offspring solutions based on the parent solutions and their objective values.
To reduce the input complexity of the LLM and mitigate the impact of numerical range on inference stability, we first normalize the inputs to the LLM. The inputs here are the parent solution set obtained in the previous step and their corresponding objective values, i.e., $w$ and their corresponding $f(w)$.
Subsequently, prompt engineering is carried out. The prompt needs to be divided into several parts:
1. A detailed description of the task;
2. The input data to be processed;
3. The expected output data format.
For example, for the optimization task in this paper, we can describe it as follows: You need to help me optimize a multi-objective optimization problem. I will provide you with multiple optimization variables and their corresponding objective values. Based on these variables and their objective values, you need to generate new offspring solutions, while ensuring that the objective values corresponding to the offspring solutions are all less than or equal to those of the parent solutions.
Next, I will provide you with the input data: $[w_1, w_2, ... w_H], [f(w_1), f(w_2), ... f(w_H)]$. Note that the output should include only the offspring solutions. Each offspring solution should start with $<$start$>$ and end with $<$end$>$. No additional explanations are needed.
With this, the prompt engineering is completed. At this point, the LLM, as a black-box operator, can generate a new round of offspring solutions.

Here is an example of prompt engineering:
\begin{tcolorbox}[colback=black!5,colframe=black,title=Example Prompt,float=htbp]
	You will assist me in minimizing a four objective task. The number of optimization variable is vector. The dimension of each variable is three. I have a set of variables along with their function values. The vector start with \text{$<$start$>$} and end with \text{$<$end$>$}. 
	
	\textbf{vector:} \text{$<$start$>$0.137,0.572,0.671$<$end$>$} \\
	\textbf{value:} \text{$<$start$>$0.025,0.034,0.041,64$<$end$>$} \\ 
	\textbf{...} \\ 
	\textbf{vector:} \text{$<$start$>$0.147,0.255,0.615$<$end$>$} \\
	\textbf{value:} \text{$<$start$>$0.017,0.022,0.047,85$<$end$>$} \\	
	Provide a new vector that different from all the vectors listed above and function values smaller than the smallest value among them. Avoid writing any code or providing explanations. Each output new vector need to begin with \text{$<$start$>$} and end with \text{$<$end$>$}.
\end{tcolorbox}

Denormalized solutions update the ideal point and optimize neighboring subproblems, the $j$ th 
Subproblem can be formulated as:
\begin{equation}
	\min_{w} \ g(\boldsymbol{w_j}|\mathbf{w_{j,i}},z) = \max_{1 \leq i \leq N+1} \left\{ \mathbf{w_{j,i}} \cdot |f_i(\boldsymbol{w_j}) - z_i| \right\},
	\label{54}
\end{equation}
where $z$ denotes the ideal point, $\mathbf{w_{j,i}}$ is the $i$th weight in $\mathbf{w_{j}}$.

Neighborhood solutions are replaced if offspring solutions exhibit superior performance on corresponding subproblems.

The Pareto solution set $\mathcal{P} = \{\boldsymbol{w_1}, \ldots, \boldsymbol{w_H}\}$ is obtained upon reaching maximum iterations. Optimal solution $w*$ is selected through:
\begin{itemize}
	\item[(1)] Filtering solutions with all objective values below predefined thresholds;
	\item[(2)] Selecting the solution with minimal $F_{age}$ from threshold-satisfying candidates.
\end{itemize}

Now, we obtain the optimal selection window size $w*$.

\subsection{Computational Complexity Analysis}

In this section, we will analyze the computational complexity of our approach.
Our computational complexity analysis refers to the standard MOEA/D. The complexity analysis can be divided into two parts: the initialization phase and the iterative phase. Since the initialization phase is executed only once, its complexity is much smaller than that of the iterative phase.

First, analyzing the initialization phase: generating $H$ weight vectors, the complexity can be expressed as $O(H)$. Next, constructing neighborhoods: For each weight vector, compute cosine similarity with all others and retrieve top $N$, the cosine similarity complexity is $O(H)$, and the top $K$ sorting is with $O(H \log K)$. Therefore, the total complexity at this point can be expressed as:
$O(H(H + H \log K)) = O(H^2 + H^2 \log K)$

Next, generating the initial solution set is with $O(H)$, computing the minimum value of each objective function over $H$ solutions is with $O(H \cdot (N+1))$. Therefore, the total complexity of the initialization phase can be expressed as:
$ O(H^2 \log K + H(N+1))$

Then entering the iterative phase: first the outer loop with $G_{max}$ generations, followed by the inner loop with H subproblems. For each subproblem, randomly selecting neighbors or global individuals is with $O(K)$, normalizing input for $P_{parents}$ (assumed size $M$), constructing prompts is with$O(M)$.
Calling the LLM for inference: assumed to be $O(C_{LLM})$. At most $3$ attempts, so the total complexity at this step is: $O(M + C_{LLM})$.
Next, updating the reference point: $O(N+1)$, iterating over K neighbors: $O(K)$, computing $g(\mathbf{o}|\mathbf{w}_j,\mathbf{z}^*)$ each time is with $O(1)$.
Therefore, the per-subproblem complexity per generation is given by:
$
O(K) + O(M +C_{LLM}) + O(N+1) + O(K) = O(M + C_{LLM} + K + N)  
$

Per-generation complexity ($H$ subproblems):
$
H\cdot O(M +C_{LLM} + K + N)  
$
Complexity over $G_{max}$ generations:
$
G{max}H\cdot O(M + C_{LLM} + K + N)  
$
Finally, archive updating (lines 20–24): iterating over H individuals: $O(H)$.

So far, the initialization Phase complexity can be described as:
$
O(H^2 \log K + H(N+1))  
$
The main Optimization Loop complexity can be described as:
$
O(G_{max}H(M +C_{LLM} + K + N)  
$
However, in practice, since the number of iterations $G_{max}$ is large, the computational complexity of the initialization phase can be ignored. Therefore, the overall algorithm complexity can be expressed as:
$
O(G_{max}H (M + C_{LLM} + K + N)  
$

\section{Numerical Simulation Results And Analysis}
\label{sec7}


This section we validate the effectiveness of the proposed framework through extensive numerical experiments. The LLM adopted in the simulations is the DeepSeek V3 model. Our baseline comparison algorithms include classical multi-objective algorithms such as NSGA-II, MEOA/D, NSGA-III, and SPEA2, as well as a deep reinforcement learning-based multi-objective algorithm (PPO-MO).
Multi-objective optimization algorithms, including MOEA/D ,NSGA-II,  NSGA-III, and SPEA2 were implemented using the pymoo framework under Python 3.9. 
All experimental results were obtained from more than 30 trials in order to eliminate occasional errors, and $K_{\text{bound}}$ was determined based on statistical results after extensive experiments.
\begin{table}[!t]
\renewcommand{\arraystretch}{1.2}
\caption{Comparison of convergence speed for Different Algorithms}
\label{tab3}
\centering
\begin{tabular}{lc}
\hline
\textbf{Algorithm} & \textbf{Running Time to Converge (s)}   \\
\hline
LLM-MOEA/D & 51.41 \\
NSGA-III & 56.22   \\
NSGA-II & 44.35  \\
MOEAD & 95.45   \\
SPEA2 & 121.32   \\
PPO-MO & 3492.79   \\
\hline
\end{tabular}
\end{table}

\begin{table}\footnotesize
	\caption{Simulation parameters}
	\label{tab4}
	\centering
	\begin{tabular}{|c|c|c|c|}
		\hline
		\textbf{Parameters} &\textbf{Value} &\textbf{Parameters} &\textbf{Value}\\
		\hline
		$N$ & $3$ & $\alpha$ & $3$ \\
		\hline
		$B$ & $20MHZ$ & $\sigma^2$ &$ 9dB$ \\
		\hline
		$v_0'$ & $20m/s$ & $v_0$ & $30m/s$ \\
		\hline
		$\mu$ & $0$ & $RRI$ & $100ms$ \\
		\hline
		$N_{SC}$ & $10$ & $N_{r}$ & 100 \\
		\hline
		$R$ & $200m$ & $N_{CA}$ & $ 10$ \\
		\hline
		$Bit$ & $500bit$ & $t_{fa}$ & $0.468ms$ \\
		\hline
		$p_{nei}$ & $0.8$ & $n_p$ & $7$ \\
		\hline
		$K$ & $20$ & $H$ & $120$ \\
		\hline
		$w^{L}$ & $20ms$ & $w^{U}$ & $150ms$ \\
		\hline
	\end{tabular}
	\vspace{-0.3cm}
\end{table}
\begin{figure}[tbp]
	\centering
	\includegraphics[width=\linewidth, scale=0.9]{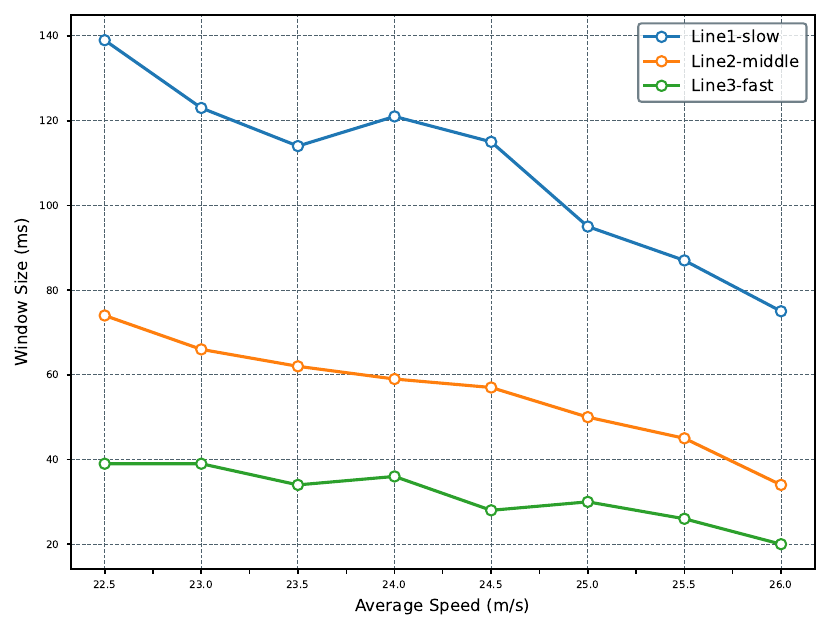}
	\caption{Selection window size VS Average velocity}
	\label{fig5}	
\end{figure}

\begin{figure}[tbp]
	\centering
	\includegraphics[width=\linewidth, scale=1.00]{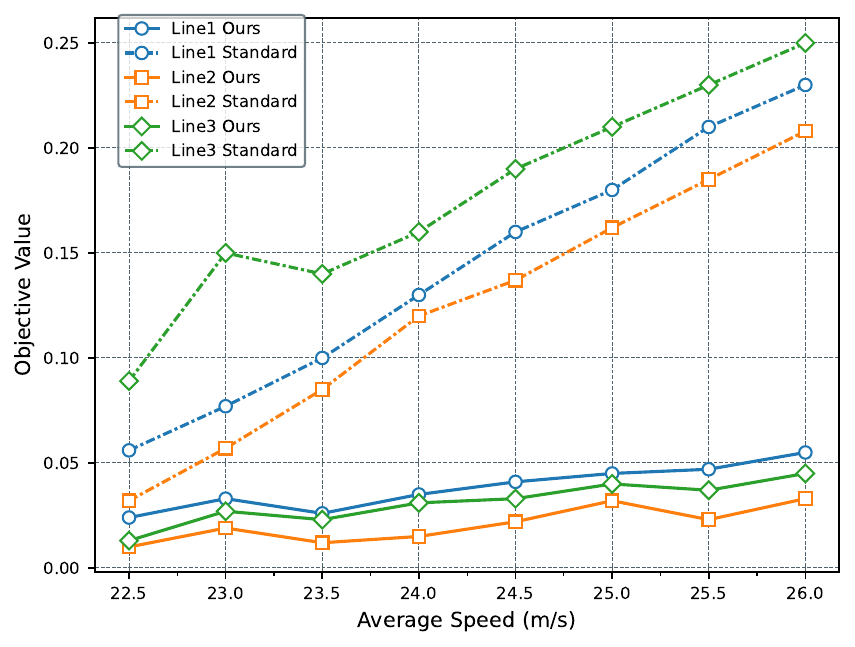}
	\caption{Objective value VS Average velocity}
	\label{fig6}	
\end{figure}
\begin{figure}[htbp]
	\centering
	\includegraphics[width=\linewidth, scale=1.00]{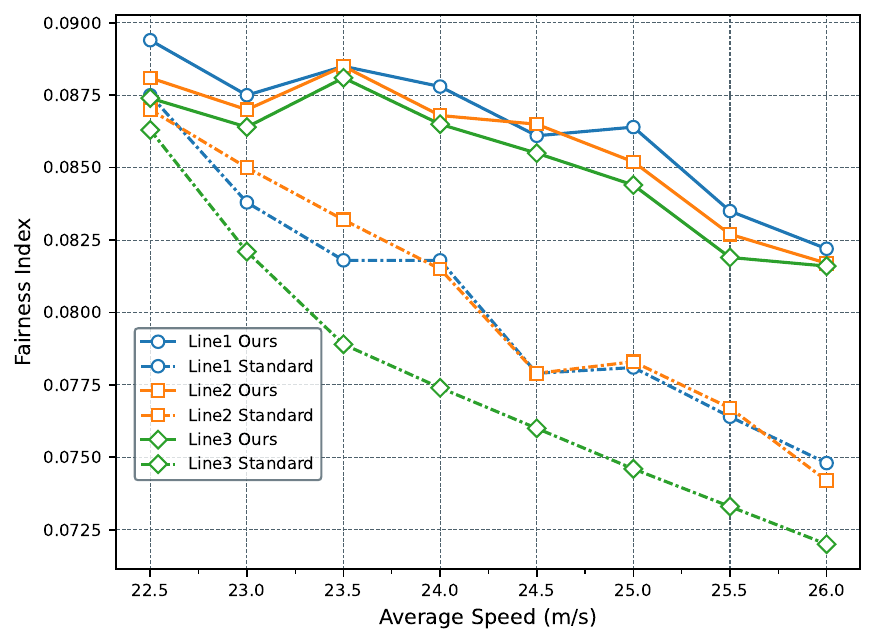}
	\caption{Fairness index VS Average velocity}
	\label{fig7}	
\end{figure}

A three-lane highway model was constructed, where vehicle speeds range between 20 m/s and 30 m/s. Speed differences across lanes are maintained at 4 m/s to simulate realistic traffic dynamics. The default selection window size and its bounds align with the 5G NR specifications.Table \ref{tab3} presents the average running time to converge of each algorithm. Although LLM has slightly longer inference time, its total runtime to reach convergence is only slightly behind NSGA-II, due to the fewer iterations required compared to other algorithms. Additional parameter configurations are summarized in Table \ref{tab4}.

Fig. \ref{fig5} illustrates the correlation between vehicle speed and selection window size, indicating a general trend of decreasing window size as average speed increases among different vehicles. This occurs because higher vehicle speeds reduce the communication duration within the RSU coverage, thereby decreasing the achievable data volume. To enhance data throughput and ensure fairness, the selection window size is dynamically reduced to minimize communication latency. Notably, faster vehicles adopt smaller windows to balance fairness across the network. We also found that sometimes when the average speed increased, the window of some vehicles increased, because the increase in the average speed was caused by other vehicles, whose speed remained the same or decreased slightly. This shows that our scheme adaptively adjusts the selection window size according to the vehicle speed.
\begin{figure}[tbp]
	\centering
	\includegraphics[width=\linewidth, scale=1.00]{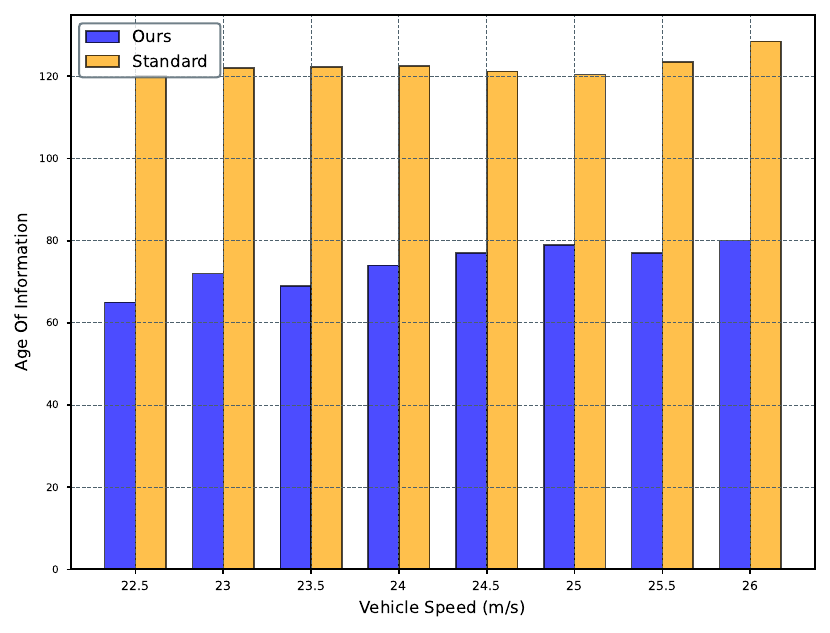}
	\caption{AoI VS Average velocity}
	\label{fig8}	
\end{figure}

\begin{figure}[htbp]
	\centering
	\includegraphics[width=\linewidth, scale=1.00]{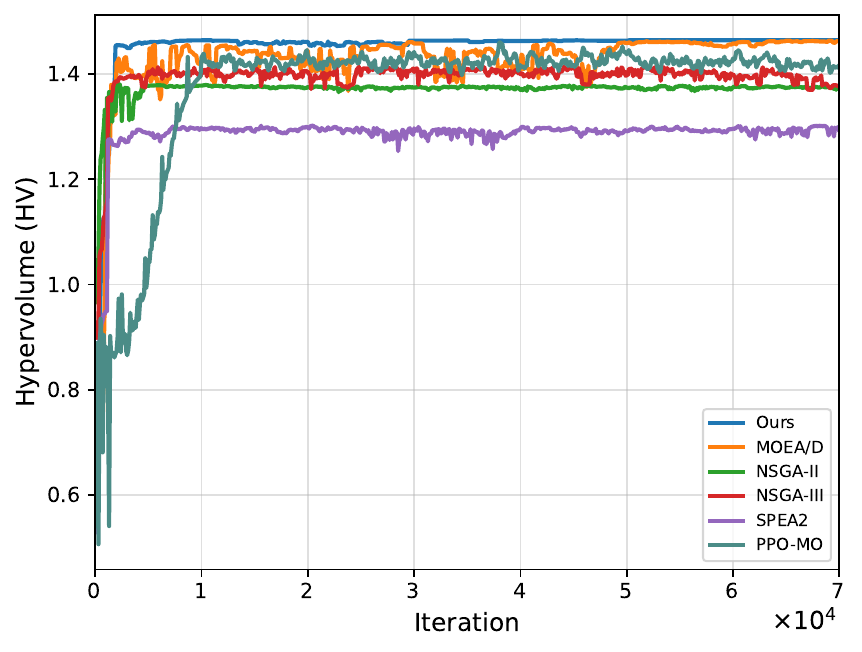}
	\caption{HV comparison}
	\label{fig9}	
\end{figure}

Fig. \ref{fig6} compares the top-$N$objective values (representing the deviation between individual vehicle fairness indices and the network average) for vehicles using standard and adaptive window strategies. As speed increases, standard-window vehicles exhibit significantly faster growth in objective values than adaptive-window vehicles. This divergence arises because fixed-window strategies fail to address the widening fairness gap between high- and low-speed vehicles. In contrast, the adaptive strategy mitigates this issue, limiting objective value growth through dynamic window adjustments.

Fig. \ref{fig7} analyzes the fairness index versus average speed. While higher network speeds degrade fairness across all vehicles, standard-window vehicles suffer severe fairness deterioration, whereas adaptive-window vehicles maintain near-stable fairness indices. The adaptive strategy compensates for speed fluctuations by optimizing window sizes, whereas fixed windows amplify speed-induced fairness variations. A more stable equity index means that vehicles with different speeds are accessing the channel and communicating with the RSU in a more equitable way.

Fig. \ref{fig8} evaluates the AoI under different strategies. Speed variations minimally impact AoI, as AoI primarily depends on window size optimization. Vehicles optimized via MOEA/D with LLM achieve lower AoI than those using fixed windows, proving the capability of the designed algorithm in minimizing the AoI.

\begin{figure}[tbp]
	\centering
	\includegraphics[width=\linewidth, scale=1.00]{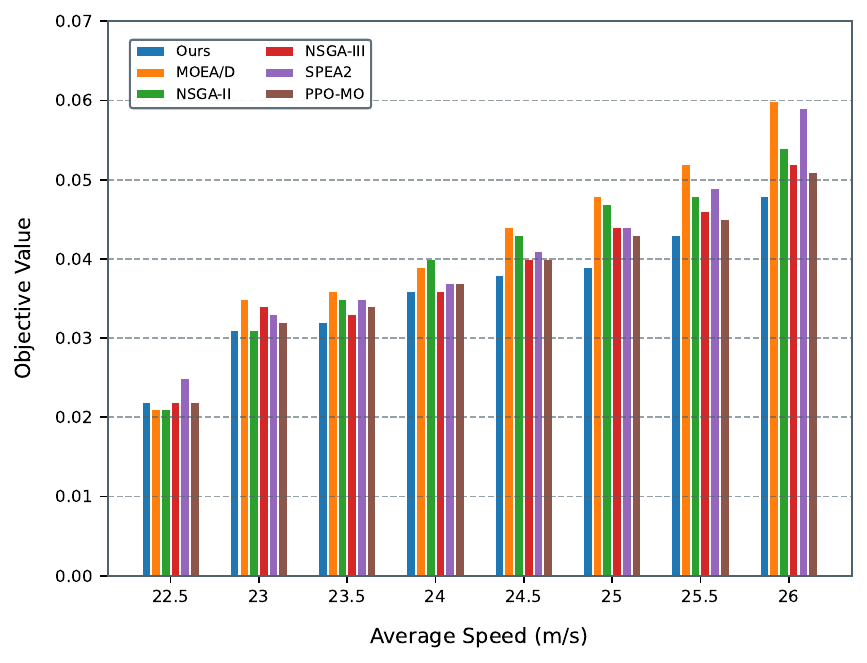}
	\caption{Objective value comparison}
	\label{fig10}	
\end{figure}
\begin{figure}[tbp]
	\centering
	\includegraphics[width=\linewidth, scale=1.00]{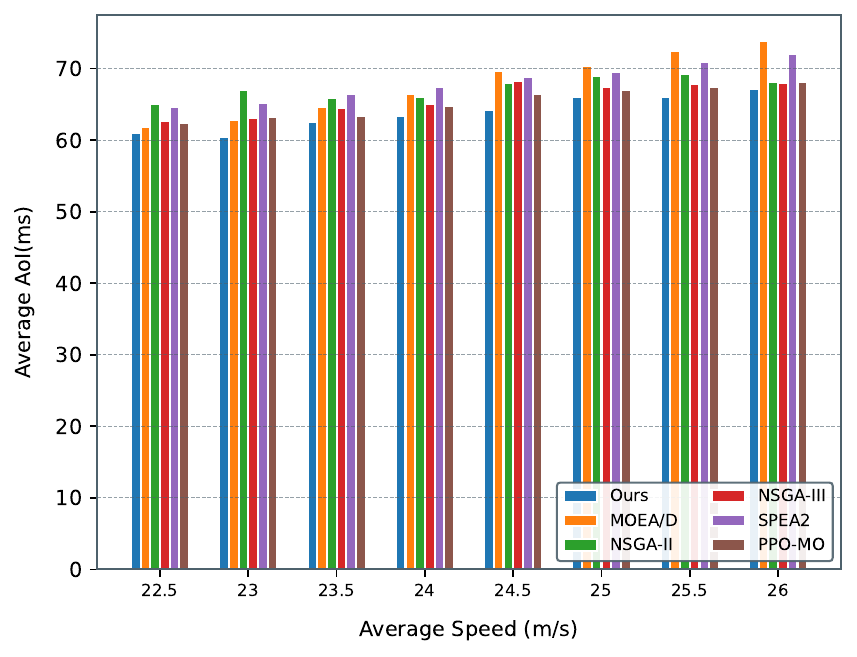}
	\caption{AoI comparison}
	\label{fig11}	
\end{figure}

Fig. \ref{fig9} to \ref{fig11} present the comparison between our proposed algorithm and other baseline algorithms. Fig. \ref{fig9} illustrates the Hypervolume (HV) comparison among different algorithms which is a commonly used metric in multi-objective optimization that evaluates the diversity, superiority, and convergence of the solution set\cite{52}. A higher HV value indicates better diversity and performance of the solution set. As shown in Fig. \ref{fig9}, the HV of the LLM-MOEA/D algorithm is the highest and achieves convergence with the fewest iterations. This indicates that LLM-MOEA/D can quickly find the optimal solution in fewer iterations, and the quality of the solution outperforms those of the other algorithms. This is attributed to the LLM-guided crossover operator, which consistently generates better offspring solutions.

Fig. \ref{fig10} shows the top-$N$ objective values of different algorithms as the average speed increases. From the figure, it can be observed that as the speed increases, the objective values of all algorithms rise, indicating that higher speeds lead to greater fairness deviations. However, although the objective values of our algorithm also increase, the growth is the smallest among them. This demonstrates that our algorithm can effectively achieve fair access under increasing speed by adjusting the selection window size.

Fig. \ref{fig11} presents the AoI performance of different algorithms as the average speed increases. As shown in the figure, the AoI of all algorithms increases slightly with speed. This is because higher speeds make it more difficult to ensure fair access, so in order to balance the joint optimization of fairness and AoI, the requirement on information freshness is relaxed. Compared with other algorithms, our algorithm achieves a lower AoI, indicating that the LLM-guided crossover operator is able to discover solution sets that Pareto-dominate those of other algorithms, thus delivering better performance in terms of AoI.

\begin{figure*}[htbp]
    \centering
    \begin{minipage}{0.3\linewidth}
        \centering
        \includegraphics[width=\linewidth]{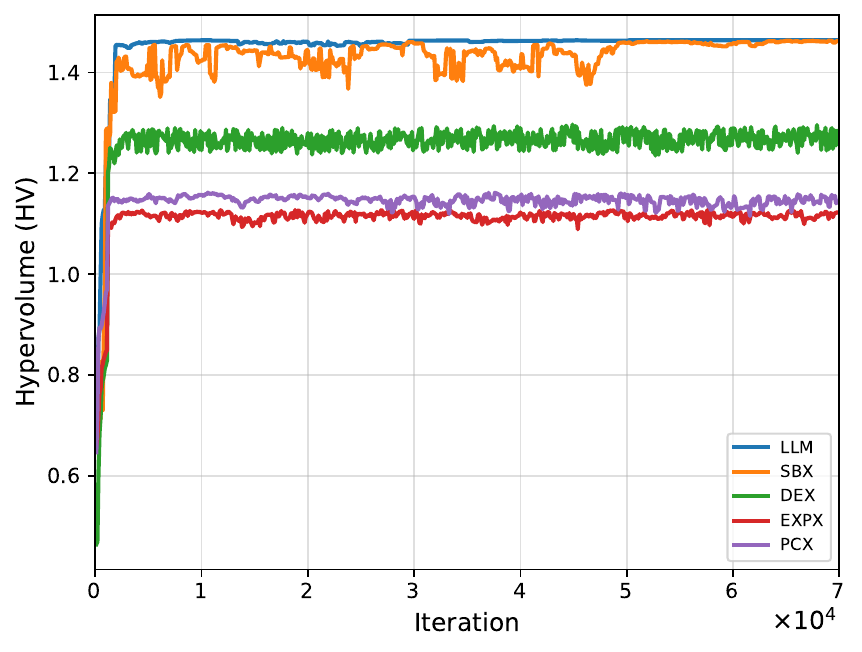}
        \caption{ Crossover HV} 
        \label{fig12}
    \end{minipage}%
    \begin{minipage}{0.3\linewidth}
        \centering
        \includegraphics[width=\linewidth]{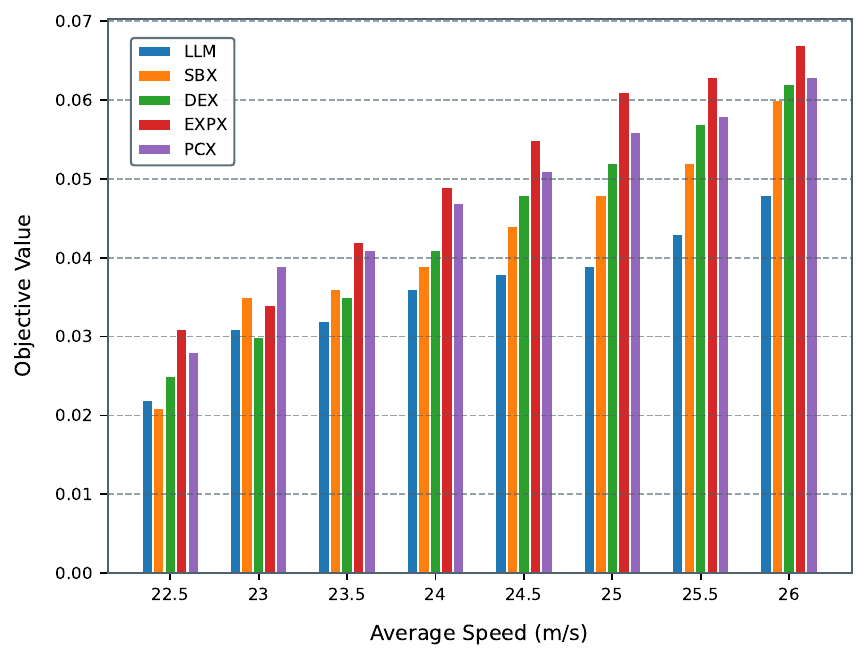}
        \caption{ Crossover objective value }
        \label{fig13}
    \end{minipage}%
    \begin{minipage}{0.3\linewidth}
        \centering
        \includegraphics[width=\linewidth]{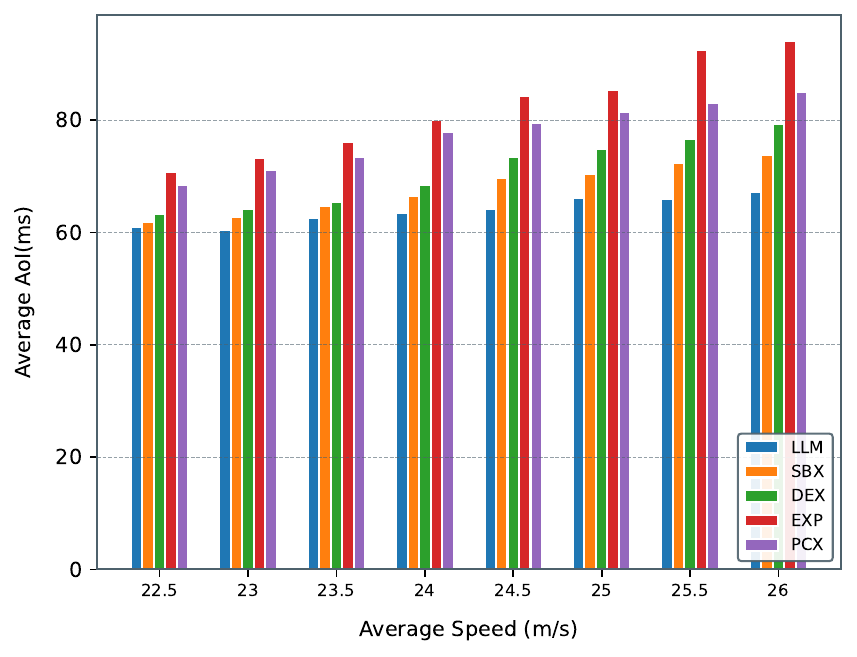}
        \caption{ Crossover AoI}
        \label{fig14}
    \end{minipage}
\end{figure*}

Fig. \ref{fig12} presents the HV convergence plot for various crossover operators within the MOEA/D algorithm framework. As shown in the figure, the crossover operator guided by LLM achieves convergence with the fewest iterations while also obtaining the highest HV value. This indicates that the LLM-guided crossover operator can provide the optimal solution in fewer iterations while maintaining solution diversity. Other crossover operators show significant fluctuations in HV values, and their HV values are consistently lower than those of the LLM-guided operator, demonstrating that LLM can deliver diverse and high-quality solutions in fewer iterations.

Fig. \ref{fig13} displays the optimization objectives of the top N solutions for each crossover operator within the MOEA/D framework. As shown in the figure, as the average speed increases, the performance of the algorithms with all crossover operators starts to decline, and the optimization objectives gradually increase. This suggests that as speed increases, it becomes more difficult for multi-objective algorithms to balance fair access and AoI. Moreover, higher speeds lead to greater fairness disparities. However, the LLM-guided crossover operator shows the lowest optimization objective value. Although fairness slightly decreases with increasing speed, the LLM-guided operator still outperforms other crossover operators, demonstrating its ability to provide the optimal solution for fair access.

Fig. \ref{fig14} illustrates the AoI performance of various crossover operators within the MOEA/D framework. As the average speed increases, the AoI of all algorithms tends to increase slightly, due to the trade-off required for fair access. However, compared to other operators, the LLM-guided crossover operator exhibits a smaller increase in AoI and its AoI value is also lower than that of the other operators. This indicates that LLM, when guiding the crossover operation, can provide more diverse and better-performing solutions while selecting those that most effectively balance fair access and AoI.

\begin{figure}[htbp]
	\centering
	\includegraphics[width=\linewidth, scale=1.00]{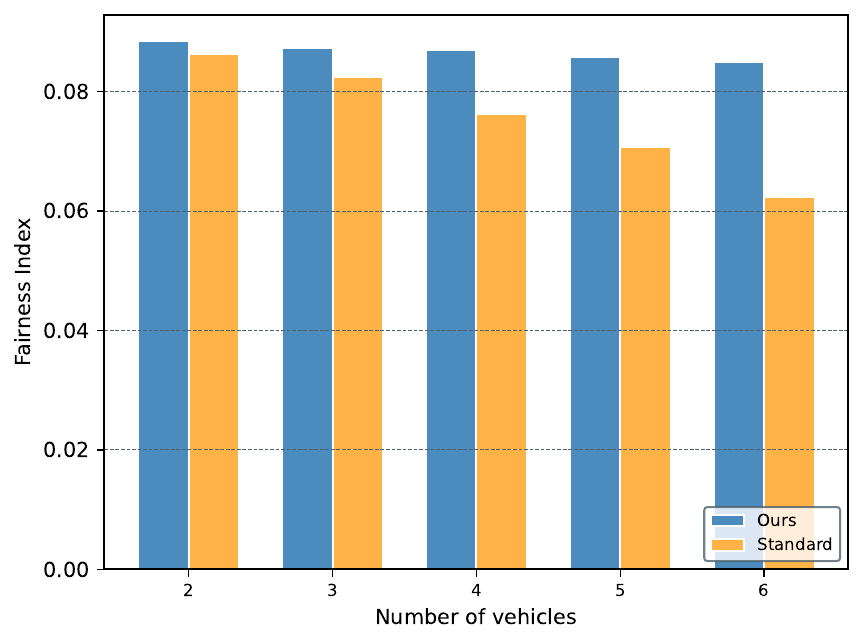}
	\caption{Objective value VS Vehicle's number}
	\label{fig15}	
\end{figure}

Fig. \ref{fig15} shows the variation of fairness index with respect to the number of vehicles under our scheme. As observed in Fig. \ref{fig15}, when the number of vehicles in the RSU increases, the fairness index of the vehicles in our scheme remains almost unchanged, whereas the fairness index of vehicles following the standard protocol decreases as the number of vehicles increases. This is because, as the number of vehicles increases, the probability of resource conflicts also increases, leading to larger differences in the amount of data transmitted by different vehicles, which results in a decline in fairness. However, our scheme can adaptively adjust the selection window size to ensure that the amount of data transmitted by each vehicle remains almost the same, thus maintaining a stable fairness index.

Fig. \ref{fig16} shows the variation of AoI with respect to the number of vehicles under our scheme. As shown in Fig. \ref{fig16}, since we measure the average AoI per vehicle, and our scheme minimizes AoI by adjusting the selection window size, the increase in the number of vehicles has minimal impact on our scheme. On the other hand, for vehicles following the standard 5G NR protocol, as the number of vehicles increases, the probability of resource conflicts also increases, leading to a significant increase in transmission time. From the figure, we can observe that the AoI of the vehicles increases, which reflects that our scheme effectively optimizes AoI even under high-pressure scenarios.

\section{Conclusion}
\label{sec8}
In this paper, we propose an enhanced SPS scheme under 5G NR V2X Mode 2. This scheme adjusts the selection window size of vehicles to eliminate unfair access issues caused by different vehicle speeds within the RSU coverage area while minimizing the average AoI, modeled using the SHS framework. We formulate a multi-objective optimization problem that jointly considers fair access and AoI minimization. To solve this problem, we employ a LLM-Based MOEA/D algorithm and determine the optimal selection window through simulations. Based on the simulation, the following major conclusions can be drawn:

\begin{figure}[tbp]
	\centering
	\includegraphics[width=\linewidth, scale=1.00]{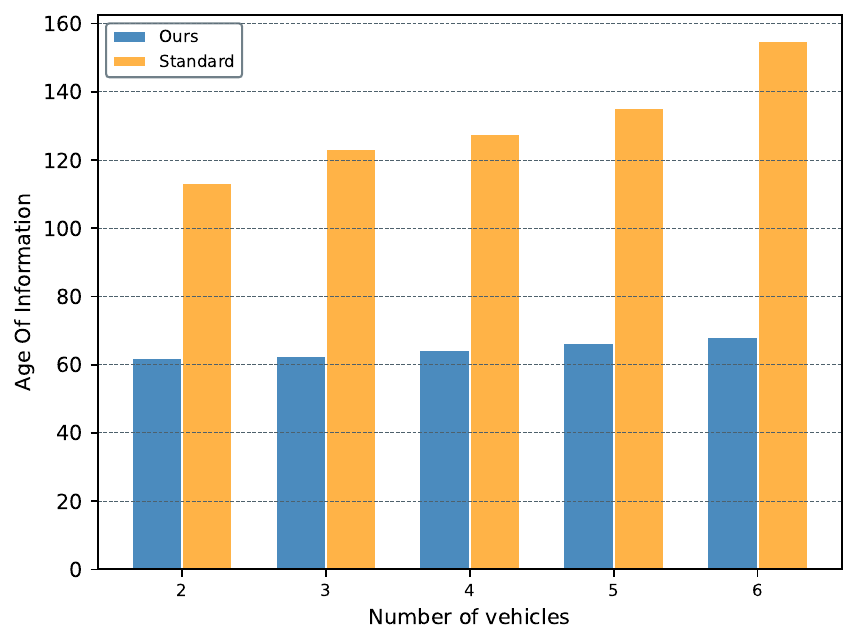}
	\caption{AoI VS Vehicle's number}
	\label{fig16}	
\end{figure}

\begin{itemize}
\item The fairness of access is strongly affected by vehicles' velocity. Higher vehicle speeds make it more challenging to achieve fairness. Therefore, a slight sacrifice in the AoI metric is necessary to maintain fairness. Achieving both optimal fairness and the lowest AoI simultaneously remains difficult.
\item AoI is a function of the selection window size, and each adjustment of the selection window primarily aims to optimize AoI. Consequently, changes in vehicle speed alone have a relatively minor impact on AoI.
\item The LLM-Based algorithm exhibits superior convergence performance compared to other algorithms. This is because large models do not generate poor solutions, ensuring that each offspring solution is Pareto-dominant.
\end{itemize}

For future work, to further optimize fairness and AoI simultaneously, additional parameters in 5G NR V2X Mode 2, such as RRI and RC size, can be explored to refine the optimization strategy.


\ifCLASSOPTIONcaptionsoff
  \newpage
\fi

\bibliographystyle{IEEEtran}

\begin{thebibliography}{1}
\bibitem{32}
A.~Camero and E.~Alba, ``Smart city and information technology: A review,''
  \emph{Cities}, vol.~93, pp. 84--94, 2019. [Online]. Available:
  \url{https://www.sciencedirect.com/science/article/pii/S0264275118304025}



\bibitem{33}
E.~Yurtsever, J.~Lambert, A.~Carballo, and K.~Takeda, ``A survey of autonomous
  driving: Common practices and emerging technologies,'' \emph{IEEE Access},
  vol.~8, pp. 58443--58469, 2020.

\bibitem{34}
W.~Xu, H.~Zhou, N.~Cheng, F.~Lyu, W.~Shi, J.~Chen, and X.~Shen, ``Internet of
  vehicles in big data era,'' \emph{IEEE/CAA Journal of Automatica Sinica},
  vol.~5, no.~1, pp. 19--35, 2018.

\bibitem{57}
X.~Wang, K.~Tao, N.~Cheng, Z.~Yin, Z.~Li, Y.~Zhang, and X.~Shen, ``Radiodiff:
  An effective generative diffusion model for sampling-free dynamic radio map
  construction,'' \emph{IEEE Transactions on Cognitive Communications and
  Networking}, vol.~11, no.~2, pp. 738--750, 2025.

\bibitem{35}
A.~Gupta and R.~K. Jha, ``A survey of {5G} network: Architecture and emerging
  technologies,'' \emph{IEEE Access}, vol.~3, pp. 1206--1232, 2015.

\bibitem{59}
J.~Shen, N.~Cheng, X.~Wang, F.~Lyu, W.~Xu, Z.~Liu, K.~Aldubaikhy, and X.~Shen,
  ``Ringsfl: An adaptive split federated learning towards taming client
  heterogeneity,'' \emph{IEEE Transactions on Mobile Computing}, vol.~23,
  no.~5, pp. 5462--5478, 2024.

\bibitem{58}
R.~Sun, N.~Cheng, C.~Li, F.~Chen, and W.~Chen, ``Knowledge-driven deep learning
  paradigms for wireless network optimization in {6G},'' \emph{IEEE Network},
  vol.~38, no.~2, pp. 70--78, 2024.
  
\bibitem{104}
Y.~Xie, Q.~Wu, P.~Fan, N.~Cheng, W.~Chen, J.~Wang, and K.~B. Letaief,
  ``Resource allocation for twin maintenance and task processing in vehicular
  edge computing network,'' \emph{IEEE Internet of Things Journal}, 2025.
  
\bibitem{111}
P.~Fan, C.~Feng, Y.~Wang, and N.~Ge, ``Investigation of the time-offset-based
  qos support with optical burst switching in wdm networks,'' in \emph{2002
  IEEE International Conference on Communications. Conference Proceedings. ICC
  2002 (Cat. No. 02CH37333)}, vol.~5.\hskip 1em plus 0.5em minus 0.4em\relax
  IEEE, 2002, pp. 2682--2686.
\bibitem{36}
C.~Gong, J.~Liu, Q.~Zhang, H.~Chen, and Z.~Gong, ``The characteristics of cloud
  computing,'' in \emph{2010 39th International Conference on Parallel
  Processing Workshops}, San Diego, CA, USA, 2010, pp. 275--279.

\bibitem{105}
Z.~Shao, Q.~Wu, P.~Fan, N.~Cheng, W.~Chen, J.~Wang, and K.~B. Letaief,
  ``Semantic-aware spectrum sharing in internet of vehicles based on deep
  reinforcement learning,'' \emph{IEEE Internet of Things Journal}, 2024.
  
\bibitem{56}
A.~Asghari and M.~K. Sohrabi, ``Server placement in mobile cloud computing: A
  comprehensive survey for edge computing, fog computing and cloudlet,''
  \emph{Computer Science Review}, vol.~51, p. 100616, 2024.

\bibitem{49}
D.~Garcia-Roger, E.~E. González, D.~Martín-Sacristán, and J.~F. Monserrat,
  ``{V2X} support in {3GPP} specifications: From {4G} to {5G} and beyond,''
  \emph{IEEE Access}, vol.~8, pp. 190946--190963, 2020.

\bibitem{50}
B.~An, J.~Lee, S.~Jang, K.~Lim, and S.~Yoon, ``Overview of {5G-NR-V2X} system
  and analysis methodology of communication performance,'' in \emph{2023 14th
  International Conference on Information and Communication Technology
  Convergence (ICTC)}, Jeju Island, Korea, 2023, pp. 1137--1142.

\bibitem{1}
M.~H.~C. Garcia, A.~Molina-Galan, M.~Boban, J.~Gozalvez, B.~Coll-Perales,
  T.~{\c{S}}ahin, and A.~Kousaridas, ``A tutorial on {5G NR V2X}
  communications,'' \emph{IEEE Communications Surveys \& Tutorials}, vol.~23,
  no.~3, pp. 1972--2026, 2021.
  \bibitem{107}
K.~Qi, Q.~Wu, P.~Fan, N.~Cheng, W.~Chen, J.~Wang, and K.~B. Letaief,
  ``Deep-reinforcement-learning-based aoi-aware resource allocation for
  ris-aided iov networks,'' \emph{IEEE Transactions on Vehicular Technology},
  2024.
\bibitem{108}
Y.~Dong, Z.~Chen, S.~Liu, P.~Fan, and K.~B. Letaief, ``Age-upon-decisions
  minimizing scheduling in internet of things: To be random or to be
  deterministic?'' \emph{IEEE Internet of Things Journal}, vol.~7, no.~2, pp.
  1081--1097, 2019.

\bibitem{109}
T.~Li, P.~Fan, Z.~Chen, and K.~B. Letaief, ``Optimum transmission policies for
  energy harvesting sensor networks powered by a mobile control center,''
  \emph{IEEE Transactions on Wireless Communications}, vol.~15, no.~9, pp.
  6132--6145, 2016.

\bibitem{46}
B.~Ko, K.~Liu, S.~H. Son, and K.-J. Park, ``Rsu-assisted adaptive scheduling
  for vehicle-to-vehicle data sharing in bidirectional road scenarios,''
  \emph{IEEE Transactions on Intelligent Transportation Systems}, vol.~22,
  no.~2, pp. 977--989, 2021.

\bibitem{60}
N.~Cheng, F.~Lyu, W.~Quan, C.~Zhou, H.~He, W.~Shi, and X.~Shen,
  ``Space/aerial-assisted computing offloading for iot applications: A
  learning-based approach,'' \emph{IEEE Journal on Selected Areas in
  Communications}, vol.~37, no.~5, pp. 1117--1129, 2019.
\bibitem{103}
X.~Gu, Q.~Wu, P.~Fan, Q.~Fan, N.~Cheng, W.~Chen, and K.~B. Letaief, ``Drl-based
  resource allocation for motion blur resistant federated self-supervised
  learning in iov,'' \emph{IEEE Internet of Things Journal}, 2024.
\bibitem{110}
Y.~Yang and P.~Fan, ``Doppler frequency offset estimation and diversity
  reception scheme of high-speed railway with multiple antennas on separated
  carriage,'' \emph{Journal of Modern Transportation}, vol.~20, no.~4, pp.
  227--233, 2012.
\bibitem{44}
R.~D. Yates, Y.~Sun, D.~R. Brown, S.~K. Kaul, E.~Modiano, and S.~Ulukus, ``Age
  of information: An introduction and survey,'' \emph{IEEE Journal on Selected
  Areas in Communications}, vol.~39, no.~5, pp. 1183--1210, 2021.
\bibitem{102}
X.~Gu, Q.~Wu, P.~Fan, N.~Cheng, W.~Chen, and K.~B. Letaief, ``Drl-based
  federated self-supervised learning for task offloading and resource
  allocation in isac-enabled vehicle edge computing,'' \emph{Digital
  Communications and Networks}, 2024.

\bibitem{61}
W.~Wang, N.~Cheng, M.~Li, T.~Yang, C.~Zhou, C.~Li, and F.~Chen, ``Value
  matters: A novel value of information-based resource scheduling method for
  cavs,'' \emph{IEEE Transactions on Vehicular Technology}, vol.~73, no.~6, pp.
  8720--8735, 2024.

\bibitem{63}
X.~Xu, Q.~Wu, P.~Fan, and K.~Wang, ``Enhanced sps velocity-adaptive scheme:
  Access fairness in {5G NR V2I} networks,'' \emph{arXiv preprint
  arXiv:2501.08037}, 2025.
  


\bibitem{64}
A.~Rolich, I.~Turcanu, and A.~Baiocchi, ``Aoi-aware and persistence-driven
  congestion control in {5G NR-V2X} sidelink communications,'' in \emph{2024 22nd
  Mediterranean Communication and Computer Networking Conference
  (MedComNet)}.\hskip 1em plus 0.5em minus 0.4em\relax IEEE,  Nice, France, 2024, pp. 1--4.

\bibitem{106}
Q.~Wu and J.~Zheng, ``Performance modeling of ieee 802.11 dcf based fair
  channel access for vehicular-to-roadside communication in a non-saturated
  state,'' in \emph{2014 IEEE International Conference on Communications
  (ICC)}.\hskip 1em plus 0.5em minus 0.4em\relax IEEE, 2014, pp. 2575--2580.

\bibitem{26}
A.~Nabil, K.~Kaur, C.~Dietrich, and V.~Marojevic, ``Performance analysis of
  sensing-based semi-persistent scheduling in {C-V2X} networks,'' in \emph{2018
  IEEE 88th Vehicular Technology Conference (VTC-Fall)}, Chicago, IL, USA, 2018, pp. 1--5.

\bibitem{27}
Y.~Feng, A.~Nirmalathas, and E.~Wong, ``A predictive semi-persistent scheduling
  scheme for low-latency applications in {LTE} and {NR} networks,'' in
  \emph{ICC 2019 - 2019 IEEE International Conference on Communications (ICC)}, Shanghai, China,
  2019, pp. 1--6.

\bibitem{28}
X.~Gu, J.~Peng, L.~Cai, Y.~Cheng, X.~Zhang, W.~Liu, and Z.~Huang,
  ``{Performance Analysis and Optimization for Semi-Persistent Scheduling in
  C-V2X},'' \emph{IEEE Transactions on Vehicular Technology}, vol.~72, no.~4,
  pp. 4628--4642, 2023.

\bibitem{29}
M.~Muhammad~Saad, M.~Ashar~Tariq, M.~Mahmudul~Islam, M.~Toaha Raza~Khan,
  J.~Seo, and D.~Kim, ``{Enhanced Semi-persistent scheduling (e-SPS) for
  Aperiodic Traffic in NR-V2X},'' in \emph{2022 International Conference on
  Artificial Intelligence in Information and Communication (ICAIIC)}, Jeju Island, South Korea, 2022, pp.
  171--175.

\bibitem{30}

S.~Daw, A.~Kar, and B.~R. Tamma, ``{On Enhancing Semi-Persistent Scheduling in
  5G NR V2X to Support Emergency Communication Services in Highly Congested
  Scenarios},'' in \emph{Proceedings of the 24th International Conference on
  Distributed Computing and Networking}, ser. ICDCN '23.\hskip 1em plus 0.5em
  minus 0.4em\relax New York, NY, USA: Association for Computing Machinery,
  2023, p. 245–253. [Online]. Available:
  \url{https://doi.org/10.1145/3571306.3571409}


\bibitem{31}

L.~Lusvarghi, A.~Molina-Galan, B.~Coll-Perales, J.~Gozalvez, and M.~L. Merani,
  ``A comparative analysis of the semi-persistent and dynamic scheduling
  schemes in {NR-V2X mode 2},'' \emph{Vehicular Communications}, vol.~42, p.
  100628, 2023. [Online]. Available:
  \url{https://www.sciencedirect.com/science/article/pii/S221420962300058X}


\bibitem{38}
P.~Park, ``Power controlled fair access protocol for wireless networked control
  systems,'' \emph{Wireless Networks}, vol.~21, pp. 1499--1516, 2015.

\bibitem{39}
W.~Zhang, X.~Wang, G.~Han, Y.~Peng, M.~Guizani, and J.~Sun, ``A load-adaptive
  fair access protocol for mac in underwater acoustic sensor networks,''
  \emph{Journal of Network and Computer Applications}, vol. 173, p. 102867,
  2021.

\bibitem{40}
J.~Gibson, G.~G. Xie, and Y.~Xiao, ``Performance limits of fair-access in
  sensor networks with linear and selected grid topologies,'' in \emph{IEEE
  GLOBECOM 2007 - IEEE Global Telecommunications Conference}, Washington, DC, USA, 2007, pp.
  688--693.

\bibitem{9}
Q.~Wu, Z.~Wan, Q.~Fan, P.~Fan, and J.~Wang, ``Velocity-adaptive access scheme
  for mec-assisted platooning networks: Access fairness via data freshness,''
  \emph{IEEE Internet of Things Journal}, vol.~9, no.~6, pp. 4229--4244, 2021.

\bibitem{41}
M.~N. Avcil, M.~Soyturk, and B.~Kantarci, ``Fair and efficient resource
  allocation via vehicle-edge cooperation in {5G-V2X} networks,''
  \emph{Vehicular Communications}, vol.~48, p. 100773, 2024. [Online].
  Available:
  \url{https://www.sciencedirect.com/science/article/pii/S2214209624000482}


\bibitem{42}
H.~Wang, J.~Xie, and M.~M.~A. Muslam, ``Fair: Towards impartial resource
  allocation for intelligent vehicles with automotive edge computing,''
  \emph{IEEE Transactions on Intelligent Vehicles}, vol.~8, no.~2, pp.
  1971--1982, 2023.

\bibitem{43}
V.~P. Harigovindan, A.~V. Babu, and L.~Jacob, ``Ensuring fair access in ieee
  802.11 p-based vehicle-to-infrastructure networks,'' \emph{EURASIP Journal on
  Wireless Communications and Networking}, vol. 2012, pp. 1--17, 2012.


  
\bibitem{101}
	L.~Cao, H.~Yin, R.~Wei, and L.~Zhang, ``Optimize semi-persistent scheduling in
	{NR-V2X}: An age-of-information perspective,'' in \emph{2022 IEEE Wireless
		Communications and Networking Conference (WCNC)}.\hskip 1em plus 0.5em minus
	0.4em\relax IEEE, Austin, TX, USA, 2022, pp. 2053--2058.
	
\bibitem{20}
M.~Azizi, F.~Zeinali, M.~R. Mili, and S.~Shokrollahi, ``Efficient aoi-aware
  resource management in {VLC-V2X} networks via multi-agent rl mechanism,''
  \emph{IEEE Transactions on Vehicular Technology}, vol.~73, no.~9, pp.
  14009--14014, 2024.

\bibitem{21}
Z.~Zhang, Q.~Wu, P.~Fan, N.~Cheng, W.~Chen, and K.~B. Letaief, ``Drl-based
  optimization for aoi and energy consumption in {C-V2X} enabled iov,''
  \emph{IEEE Transactions on Green Communications and Networking}, pp. 1--1,
  2025.

\bibitem{22}
A.~Maatouk, M.~Assaad, and A.~Ephremides, ``On the age of information in a csma
  environment,'' \emph{IEEE/ACM Transactions on Networking}, vol.~28, no.~2,
  pp. 818--831, 2020.

\bibitem{23}
R.~D. Yates and S.~K. Kaul, ``The age of information: Real-time status updating
  by multiple sources,'' \emph{IEEE Transactions on Information Theory},
  vol.~65, no.~3, pp. 1807--1827, 2019.

\bibitem{62}
Y.~Qiu, M.~Chen, H.~Huang, W.~Liang, J.~Liang, Y.~Hao, and D.~Niyato,
  ``Spotlighter: Backup age-guaranteed immersive virtual vehicle service
  provisioning in edge-enabled vehicular metaverse,'' \emph{IEEE Transactions
  on Mobile Computing}, vol.~23, no.~12, pp. 13375--13391, 2024.

\bibitem{24}
L.~Cao, H.~Yin, R.~Wei, and L.~Zhang, ``Optimize semi-persistent scheduling in
  {NR-V2X}: An age-of-information perspective,'' in \emph{2022 IEEE Wireless
  Communications and Networking Conference (WCNC)}, Austin, TX, USA, 2022, pp. 2053--2058.

\bibitem{25}
M.~M. Saad, M.~A. Tariq, J.~Seo, M.~Ajmal, and D.~Kim, ``Age-of-information
  aware intelligent mac for congestion control in {NR-V2X},'' in \emph{2023
  Fourteenth International Conference on Ubiquitous and Future Networks
  (ICUFN)}, Paris, France, 2023, pp. 265--270.

\bibitem{18}
Q.~Wu, Z.~Wan, Q.~Fan, P.~Fan, and J.~Wang, ``Velocity-adaptive access scheme
  for mec-assisted platooning networks: Access fairness via data freshness,''
  \emph{IEEE Internet of Things Journal}, vol.~9, no.~6, pp. 4229--4244, 2021.

\bibitem{12}
H.~Q. Ngo, E.~G. Larsson, and T.~L. Marzetta, ``Energy and spectral efficiency
  of very large multiuser mimo systems,'' \emph{IEEE Transactions on
  Communications}, vol.~61, no.~4, pp. 1436--1449, 2013.

\bibitem{53}
K.~E. Baddour and N.~C. Beaulieu, ``Autoregressive modeling for fading channel
  simulation,'' \emph{IEEE Transactions on Wireless Communications}, vol.~4,
  no.~4, pp. 1650--1662, 2005.

\bibitem{54}
R.~Alieiev, T.~Hehn, A.~Kwoczek, and T.~K{\"u}rner, ``Predictive communication
  and its application to vehicular environments: Doppler-shift compensation,''
  \emph{IEEE Transactions on Vehicular Technology}, vol.~67, no.~8, pp.
  7380--7393, 2018.

\bibitem{13}
C.~Brady, L.~Cao, and S.~Roy, ``Modeling of {NR C-V2X mode 2 }throughput,'' in
  \emph{2022 IEEE International Workshop Technical Committee on Communications
  Quality and Reliability (CQR)}.\hskip 1em plus 0.5em minus 0.4em\relax IEEE,
  Arlington, VA United States, 2022, pp. 19--24.

\bibitem{17}
Y.~Sun, E.~Uysal-Biyikoglu, R.~D. Yates, C.~E. Koksal, and N.~B. Shroff,
  ``Update or wait: How to keep your data fresh,'' \emph{IEEE Transactions on
  Information Theory}, vol.~63, no.~11, pp. 7492--7508, 2017.

\bibitem{19}
M.~C. Lucas-Esta{\~n}, B.~Coll-Perales, T.~Shimizu, J.~Gozalvez, T.~Higuchi,
  S.~Avedisov, O.~Altintas, and M.~Sepulcre, ``An analytical latency model and
  evaluation of the capacity of {5G NR} to support {V2X} services using {V2N2V}
  communications,'' \emph{IEEE Transactions on Vehicular Technology}, vol.~72,
  no.~2, pp. 2293--2306, 2022.

\bibitem{51}
F.~Liu, X.~Lin, S.~Yao, Z.~Wang, X.~Tong, M.~Yuan, and Q.~Zhang, ``Large
  language model for multiobjective evolutionary optimization,'' in
  \emph{International Conference on Evolutionary Multi-Criterion
  Optimization}.\hskip 1em plus 0.5em minus 0.4em\relax Springer, Canberra, ACT, Australia, 2025, pp.
  178--191.


\bibitem{55}
3GPP, ``{Release 16 Description; Summary of Rel-16 Work Items},'' {3rd
  Generation Partnership Project (3GPP)}, Technical report (TR) 21.916, Apl, 
  2020, version 16.2.0. [Online]. Available:
  \url{https://portal.3gpp.org/desktopmodules/Specifications/SpecificationDetails.aspx?specificationId=3493}


\bibitem{52}
N.~Riquelme, C.~Von~Lücken, and B.~Baran, ``Performance metrics in
  multi-objective optimization,'' in \emph{2015 Latin American Computing
  Conference (CLEI)}, Arequipa, Peru, 2015, pp. 1--11.





\end{thebibliography}

\end{document}